\documentclass[referee,useAMS,usenatbib,usegraphicx,ansinew]{mn2e}

\title[ Two-colour photometry of UU Sge and V477 Lyr]
{Two-colour photometry of the binary planetary nebula nuclei UU Sagitte and V477 Lyrae: oversized
secondaries in post-common-envelope binaries}
\author[M. Af\c{s}ar and C. \.{I}bano\v{g}lu]{M. Af\c{s}ar$^{1}$\thanks{E-mail:
melike.afsar@ege.edu.tr (MA)} and C. \.{I}bano\v{g}lu$^{1}$\\
$^{1}$Department of Astronomy and Space Sciences, Faculty of Science, Ege University, 35100  Bornova, \.{I}zmir, Turkey}

\begin{document}

\date{Accepted 2008 September 3. Received 2008 September 3; in original form 2008 April 23}

\pagerange{\pageref{firstpage}--\pageref{??}} \pubyear{2008}

\maketitle
\label{firstpage}

\begin{abstract}

We present new {\it V--\/} and {\it R--\/}passband CCD photometry of UU Sge and V477 Lyr, the eclipsing binary nuclei
of the planetary nebulae Abell 63 and Abell 46, respectively. We have performed a simultaneous analysis of {\it VR} light--curves
and estimated the effective temperatures for the primary and secondary stars to be 78 000 $\pm$ 3000 and 6136 $\pm$ 240 $\rmn{K}$ for UU Sge, 49 500 $\pm$ 4500 and 3874 $\pm$ 350 
$\rmn{K}$ for V477 Lyr. We have also reanalysed the previously measured radial velocities and combined the results with those obtained from the analysis of the light curves to derive 
absolute parameters of the components. The secondary stars have larger radii than expected from their main--sequence counterparts at the same masses. We have determined the post--common 
envelope ages and the thermal time scales of the systems and examined the possible reasons of expanded radius of the secondary components, together with some selected post-common 
envelope binaries. We conclude that
the secondary components of the nuclei of the planetary nebulae are still out of thermal equilibrium along with two post-common envelope systems: HS 1136+6646 and RE 1016-053. For other 
systems, magnetic activity has been suggested as the more plausible reason for their expanded radii. We have also estimated the common--envelope efficiency parameters of UU Sge and V477 
Lyr.        

\end{abstract}


\section{Introduction}

The classical approach to planetary nebulae (PNe) is that they are envelopes ejected by single stars (0.8$\leq$$M/M$$_{\odot}$$\leq$10) at the end of the asymptotic giant branch (AGB) 
stage. However, up to date, it has been reported that at least 10-15 per cent of planetary nebula nuclei (PNNi) are very close binaries with periods in the range of hours to a few days 
(\citealt{b11}; \citealt{b12}; \citealt{b90}). Recently, some radial velocity surveys have been carried out in order to search for new variable PNNi, which have periods too long to be 
detected using photometry (e.g. \citealt{b22}; \citealt{b2}). The existence of these objects has been explained by means of common envelope (CE) ejection by several authors 
(\citealt{b59}; \citealt{b41}; \citealt{b42}). According to CE evolution theories, the atmosphere of the primary component will expand during the red giant branch (RGB) or AGB evolution, 
and depending on the initial separation of the binary system, both components will be engulfed by extended atmosphere of the primary. The gravitational drag forces arise due to orbital 
motion of the secondary component around the primary's core (spiral-in process) will cause orbit to shrink dramatically. Because of the conservation of total angular momentum, the 
orbital angular momentum will be transferred to the CE and as a result the envelope is spun up and ejected from the system. Thus, the outcome will be a close binary as a nucleus of a 
planetary nebula. The efficiency of which orbital energy is deposited into CE ejection is defined as ``efficiency parameter", denoted as $\alpha$$_{\rmn{CE}}$. Our knowledge on the CE 
evolution is highly limited due to the difficulties in determining of stellar and nebular parameters. However, the central stars of the PNe, which are also members of eclipsing binary 
systems, are the direct evidence of the existence of the CE phase and provide an opportunity to drive accurate absolute parameters of these objects. 
	
	There are a number of these objects and only four of them are eclipsing binaries: UU Sge, V477 Lyr, BE UMa and SuWt 2. The other close binary nuclei are known to be as binaries 
either due to reflection effect (see \citealt{b23} for a detailed study) or ellipsoidal variation they show in their light curves, or they are just single-lined spectroscopic binaries.  
	
	The eclipsing binary nuclei UU Sge, the nucleus of Abell 63, and V477 Lyr, the nucleus of Abell 46, were previously studied by several authors. The eclipsing nature of UU Sge was 
first reported by Hoffleit in 1932. After three decades the presence of a low-surface brightness nebula was discovered by \citet{b1}, and \citet{b8} showed the connection between PN and 
UU Sge. The first photoelectric photometric observations of the system were made by \citet{b9}. They obtained a light curve in {\it B--\/} passband and indicated that the system has a 
total eclipse at the primary conjunction. They also noted the extreme heating effect in the system due to large temperature difference between the components. From the solution of the 
light curve they concluded that the system contains an sdO primary of $\sim$0.9 M$_{\odot}$ with a temperature of about 35 000 $\rmn{K}$ and a dK secondary of $\sim$0.7 M$_{\odot}$. They 
also emphasized that the observations of the system was contaminated by the nearby optical companion $\sim$4{\arcsec} (later to be measured as $\sim$2{\arcsec}.8 by \citealt{b77}) to the 
east of UU Sge and they could not determine the true depth of primary minimum accurately. However, they estimated the contamination of the optical companion from the primary minimum 
observations and suggested that the actual depth of the primary eclipse should be about 4.3 mag. 
	
	Later, \citet{b16} obtained the {\it V--\/}passband light curve of the system. They found the O-type primary component has a temperature of not much higher than 30 000 $\rmn{K}$ and 
the secondary component is around 6000 $\rmn{K}$. \citet{b74} discovered the soft X--ray emission feature of UU Sge and suggested that the observed X--ray originates in the corona of the 
late-type star.
	
	\citet{b77} made spectroscopic observations of the system at several phases. Using the spectrum taken during the primary minimum they estimated the spectral type of the secondary 
component as G7 V. From the spectra obtained at out-of-eclipse they deduced that the exciting hot central star is an sdO type PN nucleus with a temperature of around 50 000 $\rmn{K}$.
	
	The first {\it V--\/} and {\it I--\/}passband CCD photometry was made by \citet{b60}. However, in {\it I--\/}passband, they only obtained the primary minimum since they aimed to use 
the primary eclipse to determine an accurate temperature for the secondary. Using the medium--resolution spectroscopy, they obtained the first radial velocities for the system. They used 
the results deduced from the analysis of radial velocities for the solution of the {\it V--\/}passband light curve. Since CCD observations enabled to subtract the contamination of the 
neighboring field star, they derived the more accurate absolute parameters for the individual components. After performing a series of light curve solution, they estimated an effective 
temperature of 117 500 $\rmn{K}$ for the primary component and fixing this value for the final light curve solution they suggested a temperature of $\sim$7300 $\rmn{K}$ for the secondary 
component.
	
	In a later study, \citet{b5} confirmed that the primary eclipse is total with a duration of $\sim$14 min. They also obtained the spectroscopic observations of the system during 
primary minima and extracted the spectrum of secondary component. Applying \citet{b47} atmosphere models to the spectra they determined a secondary component's temperature of 
6250$\pm$250 $\rmn{K}$. Adopting this value for the secondary component temperature they revised the preliminary light curve analysis. Solutions indicated a primary component with a 
temperature of 87 000$\pm$13 000 $\rmn{K}$ for a limb-brightened (\citealt{b11}) secondary component. 
	
	V477 Lyr, the eclipsing binary nucleus of the PN Abell 46, has a similar light curve to UU Sge, except that the eclipse is partial (\citealt{b10}). The system shows a very large 
reflection effect which creates sine--like distortion on the light curve with an amplitude of $\sim$0.6 mag.
	
	Although its eclipsing nature was reported more than two decades ago, a few light curve solutions of V477 Lyr have been published. \citet{b63} presented a solution based on a light 
curve analysis by L.W. Twigg (communicated by Bond, 1981). This solution yielded an orbital inclination of 80$\fdg$8 with a mass ratio of $M_{1}$/$M_{2}$$\sim$2.3, radii of 
$R_{1}$$\sim$0.09 R$_{\odot}$ and $R_{2}$$\sim$0.20 R$_{\odot}$. The temperature of the primary component was taken as $T_{1}$$\sim$105 000 $\rmn{K}$. Considering a typical mass value 
for the central star of a planetary nebula (\citealt{b67}), Ritter assumed a mass for the primary component as $M_{1}$$\sim$0.6 M$_{\odot}$, which yields a secondary component mass of 
$M_{2}$$\sim$0.25 M$_{\odot}$. Also, \citet{b11} cited the results of a preliminary analysis by J. Kaluzny and a temperature of 60 000 $\rmn{K}$ for the primary component was suggested.
	
	The most recent and detailed solutions of both light and radial velocity curves were made by \citet{b61}. They obtained the medium-resolution spectra and a new {\it V--\/}passband 
CCD photometry of the system and presented the first complete light curve of V477 Lyr. They performed a series of light curve solution for the effective temperature of primary component 
ranging from 30 000 to 130 000 $\rmn{K}$. Using the mass ratio ($M_{2}$/$M_{1}$=0.29) obtained from the radial velocities and adopting the limb-brightening for the secondary component, 
they achieved a best fit to the observations for the effective temperature of 60 000 $\rmn{K}$. Thus, the {\it V--\/}passband light curve solution yielded an inclination of 
$\sim$80$\fdg$5, and a temperature of $\sim$5300 $\rmn{K}$ for the secondary component.

	In this paper, we present new two--colour CCD photometry of UU Sge and V477 Lyr. The aim is to constrain the effective temperatures of the components through simultaneous solutions 
of the two--colour light curves. We will estimate the common envelope efficiencies, $\alpha$$_{\rmn{CE}}$, and discuss reasons for the oversized secondaries that we find.    

\section[]{OBSERVATIONS AND DATA REDUCTION}
\subsection{UU Sge}

	We have obtained the Johnson {\it V--\/} and {\it R--\/}passband photometry of the system at the National Observatory of the Scientific and Technological Research Council of Turkey 
(T\"{U}B\.{I}TAK--TUG) in Antalya. TUG is established on the mountain of Bak{\i}rl{\i}tepe (in the West Taurus Mountain ranges), at a height of about 2500 m. Observations were carried 
out using a 1.5-m Cassegrain telescope (RTT 150) equipped with 2048${\times}$2048 back illuminated CCD camera which has an area of 9${\arcmin}$.1${\times}$9${\arcmin}$.1 and a scale of 
0.26${\arcsec}$/pixel. We observed the system on the nights between July 4 and 7 and on the nights of 2002 August 6, 7 and 11. Since the orbital period of UU Sge is relatively short 
($\sim$11.5 hours) and the system has a very deep primary minimum ($\sim$4.5 mag), exposure times were limited between 60 and 120 s. The light curves were phased by using the ephemeris 
given by \citet{b44},
\[
 \rmn{HJD \, (Min.I)}=2451766.5285+0.46506921{\it E}\,\rmn{d}.
\]

	UU Sge has a nearby optical companion $\sim$2${\arcsec}$.8 to the east of the binary nucleus. The visual magnitude of the companion is V=15.87 mag. (\citealt{b19}). The visual 
magnitudes of the binary system itself at maxima and in mid--eclipse are V$_{max}$=14.67 mag and V$_{min}$=19.24 mag (PB93), respectively.
	
	The data were reduced using the appropriate tasks in {\sevensize IRAF}\footnote{{\sevensize IRAF} is distributed by the National Optical
Astronomy Observatories, which are operated by the Association of Universities for Research in Astronomy, Inc., under cooperative agreement with
the National Science Foundation.}. First a master bias was created and subtracted from the images. Then we used an averaged master flat-field, which we created using the flat frames 
taken both during the evening and morning twilight, and normalized all the images. Since UU Sge is located within a crowded region, images were also trimmed right after bias and 
flat-field corrections. After first reduction steps were completed, {\sevensize IRAF/DAOPHOT} package was used to apply the crowded field photometry. Thereby, we were able to make 
profile fitting and PSF (Point Spread Function) modeling of the stars in the target field and subtract the light contamination of the nearby optical companion. We measured the PSF 
magnitudes of UU Sge and five nearby stars to apply differential photometry. These five stars were chosen as comparison star candidates which were previously determined by PB93 (C1, C2, 
C3, C5 and C6: Fig.1, therein). After investigating all comparison candidates for a possible variability, we decided to use C3 as the comparison, and we determined the differential 
magnitudes for UU Sge as plotted in Fig.1. 
	
	We have determined the depth of the total eclipse to be 4.450$\pm$0.075 and 3.764$\pm$0.074 mag for {\it V--\/} and {\it R--\/}passband, respectively. The depth for the {\it 
V\/}--passband is somewhat larger than that of BPH94. The time difference between the fourth and first-contact points was determined as $\sim$1.26 hour, and the duration of the total 
eclipse at the bottom of the primary minimum was estimated as $\sim$13.4 min. Not only the short time length but also the integration time increase at the bottom of the total eclipse, 
with a relatively long uploading time of the camera, brought about to obtain only a few data points in conjunction. Moreover, the minimum brightness of UU Sge is close to the magnitude 
limit (in case of photometry) for a 1.5-m telescope, which led to have scattered data points at the primary conjunction with errors around as much as 0.06 mag.
\begin{figure}
\center
 \includegraphics[width=120mm]{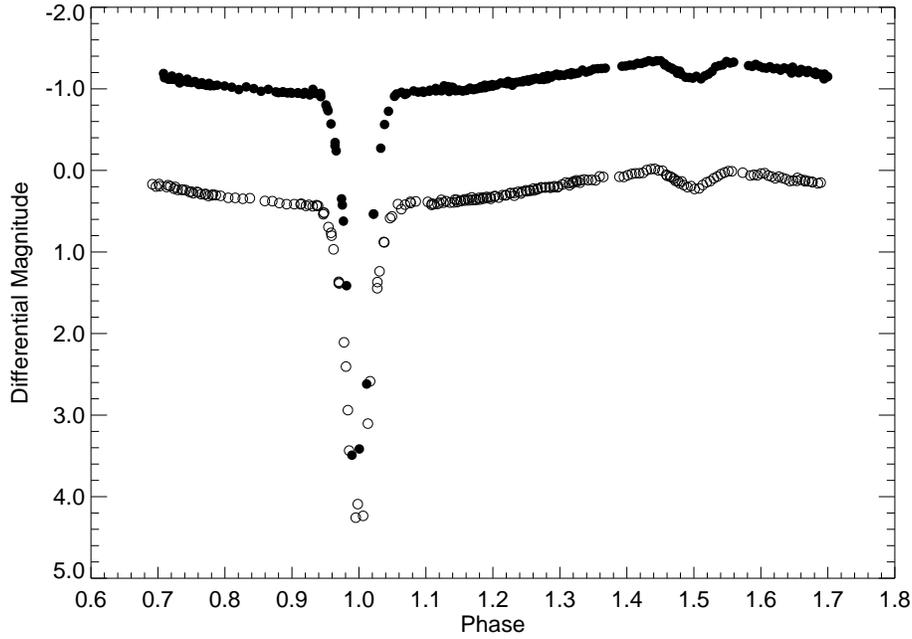}
 \caption{{\it V--\/} and {\it R--\/}passband light curves of UU Sge (filled and open circles, respectively).}
\end{figure}

\subsection{V477 Lyr}

	Two--passband (Johnson {\it V\/} and {\it R\/}) CCD photometric observations of V477 Lyr were carried out using the same setup as described for UU Sge. We observed the system during 
the nights from 2002 August 6 to 10. V477 Lyr has a partial eclipse at the primary minimum and it is not as deep as that of UU Sge. We set the exposure time as 60 s during the overall 
phases. The ephemeris used to phase the data was taken from \citet{b44},

\[
\rmn{HJD \, (Min.I)} = 2451766.4038 + 0.47172909{\it E}\,\rmn{d}.      
\]

After bias subtraction and flat-fielding, the data reduction was performed applying the aperture photometry package of {\sevensize IRAF/APPHOT}. The magnitudes of the stars including 
V477 Lyr and other five comparison candidates, which were previously found to be constant by PB94 (C1, C3, C4, C5 and C6: Fig.1, therein), were measured. The errors of the magnitude 
measurements at primary minima did not exceed 0.018 and 0.010 mag in {\it V--\/} and {\it R--\/}passband, respectively. The differential magnitudes were computed with respect to the 
comparison star C6. The light curves of the system obtained in {\it V--\/} and {\it R--\/}passband are shown in Fig.2.

\begin{figure}
\center
 \includegraphics[width=120mm]{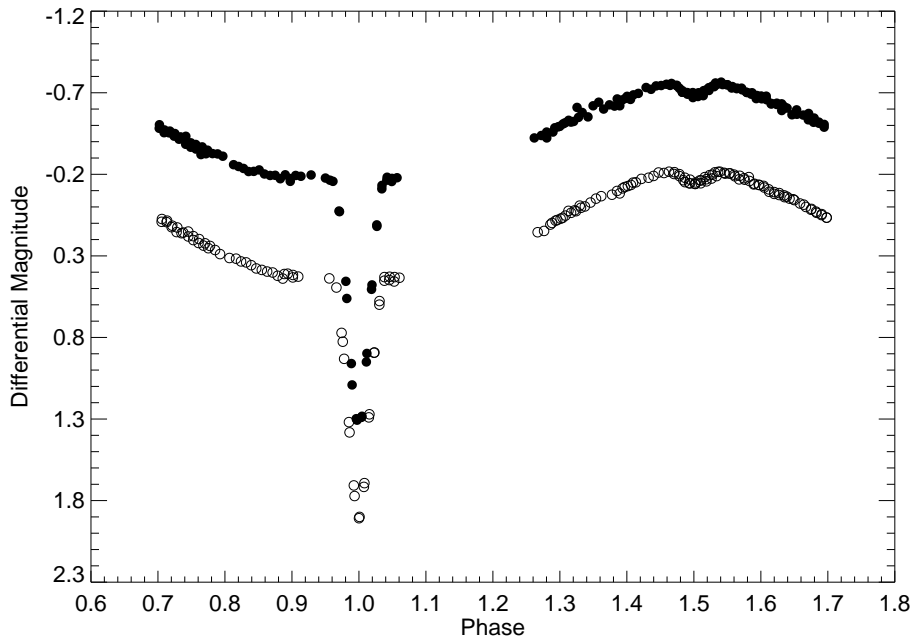}
 \caption{{\it V--\/} and {\it R--\/}passband light curves of V477 (filled and open circles, respectively).}
\end{figure}

\section{Reanalysis of the Radial Velocities}

	The first radial velocities of UU Sge and V477 Lyr were previously obtained and published by PB93 and PB94, respectively. 
By using both absorption and emission lines (which move in anti--phase with the absorption lines and thought to originate on or very close to the surface of the secondary component) they 
measured the radial velocities of the systems. They used one of the spectra of the related object (UU Sge or V477 Lyr) as a template to calculate the cross--correlations. Therefore, as 
mentioned by the authors, only the estimates of the velocity semi--amplitudes are meaningful while the systematic velocities of the systems remain unknown. That is why we were not be 
able to make simultaneous light and velocity curve solutions. However, we reanalysed the radial velocities by using the corrected radial velocity data given by PB93 and PB94 in order to 
increase the precision of the parameters and used them as inputs during the simultaneous solution of the two--colour light curves. The revised results of the radial velocity curve 
solutions are given in Table 1. The parameters $K_{1}$ and $K_{2}$ stand for the velocity semi--amplitudes of the primary and secondary components, $q$ is the mass ratio 
($M_{2}$/$M_{1}$) and $a$ is the projected semimajor axis of the related system's orbit.      

\begin{table*}
\centering
\caption{Results of the radial velocity analysis.}
 \begin{tabular}{@{}l|c|c|c|c@{}} 
   \hline
   \hline
Targets & $K_{1}$ & $K_{2}$ & $q$ & $asini$ \\
& $_{(km/s)}$ & $_{(km/s)}$ &  & $_{(R_{\odot})}$\\
\hline
UU Sge &   83.8	&	182.8	& 	0.458	&	2.45 \\
& $\pm$6.7	&	$\pm$7.4	& 	$\pm$0.034	&	$\pm$0.09 \\
\hline
V477 Lyr	&51.9	&	182.1	& 	0.285	&	2.18 \\
& $\pm$6.4	&	$\pm$7.1	& 	$\pm$0.026	&	$\pm$0.09\\
  \hline
 \end{tabular}
 \end{table*}

\section{Simultaneous Light Curve Solutions}
\subsection{UU Sge}

	We have performed a simultaneous solution of the {\it V--\/} and {\it R--\/}passband light curves by employing the updated (2003) version of Wilson-Devinney (\citealt{b80}, 
\citealt{b78}, 1990, 2003) code (hereafter W--D). This program is written for the analysis of light and velocity curves of binary stars which have either circular or eccentric orbits. 
The program has different `modes' in terms of Roche lobe filling conditions of the components. Examination of the light curve of UU Sge revealed a detached system and a circular orbit as 
expected for such a close binary. Therefore, we used `Mode 2' which is developed for detached binaries. Since the reflection effect is highly strong as seen in the light curves, a 
detailed treatment to the reflection effect was applied during the solutions.
	
	The close binary systems, such as PNNi or post-CE (post-common envelope) binaries (subdwarf or white dwarf (WD) primary + low mass secondary) have large effective temperature 
differences between the components and show extreme reflection effects in their light curves due to irradiation around the substellar points of the cool secondaries. The modeling of the 
light curves of such systems is quite difficult and requires special treatment. As discussed by \citet{b33}, the temperature on the irradiated hemisphere of the cool secondary star is 
highly enhanced, therefore, contrary to the expected values ($A_{2}$=0.5, as generally assumed for cool stars) an albedo around or greater than one ($A_{2}$$\geq$1) is needed for 
securing a good fit to the detailed shape of the reflection effect (see also \citealt{b85}). Previous studies also showed that the negative values of the limb-darkening coefficients are 
needed (limb-brightening, e.g. PB94, BPH94) due to same reason of obtaining more precise modeling of the light curves. In fact, by adopting negative limb-darkening coefficients for the 
secondary components of UU Sge and V477 Lyr, BPH94 and PB94 achieved more reasonable light curve solutions for albedo values greater than one.

	We have performed a number of solutions by taking into account the previous approaches to the light curve modeling of the similar systems. The mass ratio `$q$' obtained from the 
radial velocity analysis, the gravity brightening exponents of the components ($g_{1}$=1.0 and $g_{2}$=0.32), and the bolometric albedo of the primary component ($A_{1}$=1) were held 
fixed in the simultaneous solutions. Our first attempt to solve the light curves by using the square--root limb--darkening coefficients did not yield a rational solution, thence we 
decided to use linear limb--darkening law. The wavelength-specific linear limb--darkening coefficients of the hot companion were adopted as 0.10 and 0.07 for $x_{1}$(V) and $x_{1}$(R), 
respectively. These values were assumed using the tables given by \citet{b24} and \citet{b84}. Linear bolometric limb--darkening coefficients for both components were adopted from 
\citet{b75}. Kurucz atmosphere models (\citealt{b49}) were used for both stars during the solutions. The adjusted parameters were the orbital inclination ($i$), the mean effective 
temperature of the secondary, ($T_{2}$--not including the effects that may influence the shape of the light curve such as reflection or spots), surface potentials of the components 
($\Omega_{1}$, $\Omega_{2}$), bandpass luminosity of the primary component ($L_{1}$), bolometric albedo of the secondary component ($A_{2}$), and the wavelength-specific linear 
limb--darkening coefficients of the secondary component ($x_{2}$(V), $x_{2}$(R) : the input values were taken from \citet{b20}). The main criterion to secure a reasonable solution was to 
achieve a physically realistic value for the albedo of the secondary component (0.5$\leq$A$_{2}$$\leq$1.0). This criterion allowed us to set a lower limit for the temperature of the 
primary component, $T_{1}$, and solution trials were started with a temperature of $T_{1}$=73 000 $\rmn{K}$. So as to find the best solution, trials were made between the temperatures of 
73 000 and 85 000 $\rmn{K}$. The temperature of the secondary component was first adopted as 3500 $\rmn{K}$ considering the expected value of such a low mass star. Iterations were 
carried out until the convergence was reached at the smallest value for the residual sum of squares ({\it rss\/}). Although the solution trials have yielded parameters that are capable 
of modeling the light curves almost equal in appearance, we have achieved the best fit around the secondary minimum for a primary temperature of 78 000 $\rmn{K}$ and a secondary 
temperature of 6136$\pm$80 $\rmn{K}$, where the {\it rss\/} value was at a minimum among other temperatures adopted in the solutions. Examination of $T_{1}$ -- {\it rss\/} variation 
yielded an approximate error of 3000 $\rmn{K}$ for the hot component's temperature. The uncertainty in the temperature of the secondary component, 80 $\rmn{K}$,  is the formal 1$\sigma$ 
error from the solution. The corrected error can be estimated as 240 $\rmn{K}$ based on the uncertainty of 3000 K in the effective temperature of the primary. The results of the best 
solution are given in Table 2 (adjusted parameters have standard errors provided by the W--D code). The light curves and the computed fits are plotted in Fig.3. Some of the astrophysical 
parameters for the components were calculated and are provided in Table 3.
\begin{table*}
\centering
\caption{Parameters obtained from the simultaneous light curve analysis of UU Sge and V477 Lyr.}
 \begin{tabular}{@{}lcc@{}}
   \hline
   \hline
Parameter&UU Sge&V477 Lyr\\
\hline

$i$ ($\degr$)	&	87.12	$\pm$	0.19	&	80.33	$\pm$	0.06\\
$\Omega_{1}$	&	7.497	$\pm$	0.040	&	13.123	$\pm$	0.099\\
$\Omega_{2}$	&	3.309	$\pm$	0.009	&	2.738	$\pm$	0.006\\
$T_{1}$ ($\rmn{K}$)	&	78 000 	&	49 500		\\
$T_{2}$ ($\rmn{K}$)	&	6136	$\pm$	80	&	3874	$\pm$	120\\
$L_{1}/(L_{1}+L_{2})_{V}$	&	0.9837	$\pm$	0.0035	&	0.9940	$\pm$	0.0029\\
$L_{1}/(L_{1}+L_{2})_{R}$	&	0.9673	$\pm$	0.0026	&	0.9811	$\pm$	0.0031\\
$A_{1}$	&	1.0		&	1.0		\\
$A_{2}$	&	0.927	$\pm$	0.037	&	0.998	$\pm$	0.036\\
$g_{1}$	&	1.0			&	1.0		\\
$g_{2}$	&	0.32		&	0.32	\\
$\langle r_{1} \rangle$	&	0.1423	$\pm$	0.0013	&	0.0779	$\pm$	0.0007\\
$\langle r_{2} \rangle$	&	0.2277	$\pm$	0.0010	&	0.2091	$\pm$	0.0003\\
$x_{1}$(bol)	&	0.616		&	0.616		\\
$x_{2}$(bol)	&	0.493			&	0.384		\\
$x_{1}$(V)	&	0.100		&	0.192	\\
$x_{1}$(R)	&	0.070		&	0.162	\\
$x_{2}$(V)	&	0.478	$\pm$	0.078	&	0.210	$\pm$	0.051\\
$x_{2}$(R)	&	-0.148	$\pm$	0.074	&	-0.121	$\pm$	0.053\\
{\it rss\/}	&	0.007		&	0.023	\\

  \hline
 \end{tabular}
 \end{table*}
 
\begin{figure}
\center
 \includegraphics[width=120mm]{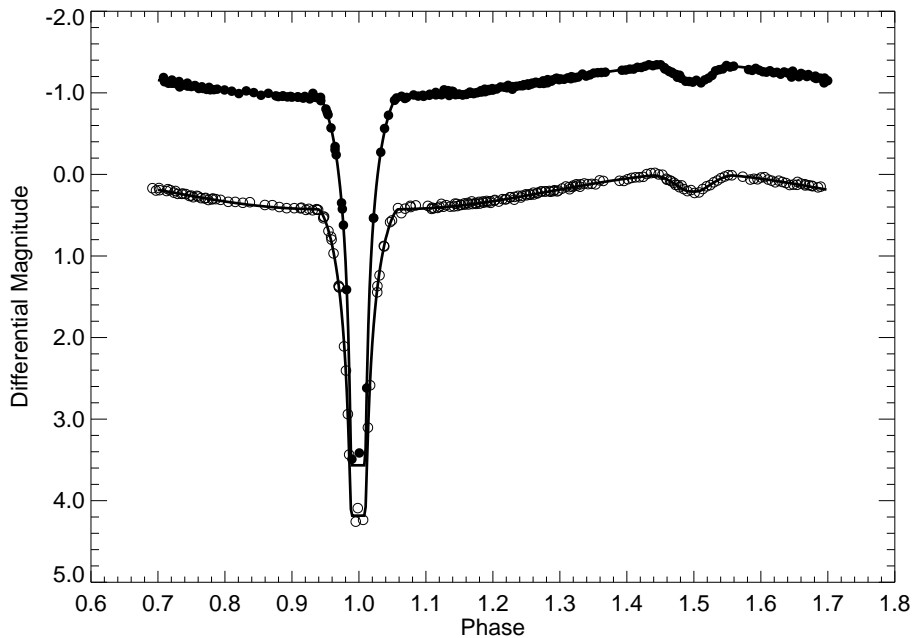}
 \caption{{\it V--\/} and {\it R--\/}passband light curves of UU Sge with computed fits.}
\end{figure}

	The mass of the secondary component seems considerably low compared to the mass of a main--sequence star with the same temperature. If we assume that the mass of the secondary 
component has not changed dramatically by the end of the CE interaction and it is just a low--mass main--sequence star, 
then the temperature seems significantly higher than that expected from an M type dwarf. The possible reasons which could explain this contradiction are discussed in $\S$ 6.      
	
	Since the simultaneous two--colour light curve solution gives us an opportunity to constrain the temperatures of the components, we believe that the temperatures derived in this 
paper are more accurate than the temperatures reported in previous studies. For example, \citet{b77} suggested a primary component temperature of 50 000 $\rmn{K}$ by comparing the merged 
ultraviolet spectrum with \citet{b48} LTE model atmospheres. They also obtained the spectrum of the secondary component during the primary eclipse (with some contamination from the sdO 
star) and assuming that the secondary is a main sequence star they classified it as G7 with an uncertainty of three spectral subclasses. BPH94 also took some spectra of the cool 
component during the primary minima. They used {\sevensize ATLAS6} computer code (\citealt{b47}) and generated several atmosphere models to find the best fit to the spectrum of the 
secondary. Under the assumption that the secondary is a main sequence star they determined a temperature of 6250$\pm$250 $\rmn{K}$. They used this value to reanalyse the light curve of 
UU Sge obtained by PB93 and gave two different results for the primary component temperature as 57 000$\pm$8000 and 87 000$\pm$13 000 K, in terms of limb-darkening and brightening 
approximation, respectively.
	
	The limb--darkening coefficients of the secondary component obtained from the solution significantly differ from those given by \citet{b20}. If we compare our values with published 
ones, we see that the {\it V--\/}passband linear limb--darkening coefficient corresponds to a temperature of $\sim$10 000 $\rmn{K}$. The {\it R--\/}passband linear limb--darkening 
coefficient has a negative value and has no reference in the published tables. The only explanation we could make for such limb--darkening coefficients would be the reflection effect 
emanated from the overheated face of the secondary, which gives rise to the ``limb--brightening effect". 

	There are several studies concerning the limb--darkening coefficients of illuminated atmospheres (e.g. \citealt{b87}, \citealt{b86}, \citealt{b88}). According to \citet{b88}, as the 
illumination flux increases, the temperature distribution in the atmosphere of the secondary component becomes more homogeneous, in other words, center-to-limb brightness variation or 
center-to-limb-darkening over the stellar disk is removed. The observed increase in the intensity of the brightness of the secondary star from its center to its limb may be explained in 
terms of horizontal energy transport (\citealt{b11}). On the other hand, the limb--darkening/brightening coefficients for the systems that have strongly irradiated atmospheres still need 
further investigation.      
	
\subsection{V477 Lyr}
	We have analysed the {\it V--\/} and {\it R--\/}passband light curves of V477 Lyr using the same code as described in the previous section. A series of simultaneous solution trials 
for a large temperature range of the primary component were performed. In order to obtain the best and realistic solution, we took into account the same criterion for the albedo of the 
secondary component, as we did for UU Sge. This criterion enable us to adopt a starting point for the temperature of hot component as 45 000 $\rmn{K}$. The fixed and adjusted parameters 
were the same as described for UU Sge. Similarly, we first attempted to apply the square root limb--darkening law during the solutions, as suggested by several authors for the stars have 
temperatures higher than 8000 $\rmn{K}$ (e.g. \citealt{b20}), but we could not reach a rational solution and decided to apply the linear limb--darkening law. The input linear 
limb--darkening coefficients for the primary and secondary components were taken from the tables published by \citet{b24}, \citet{b84} and \citet{b20}. Solution trials were made for the 
temperatures between 45 000 and 80 000 $\rmn{K}$. The best solution was achieved for the temperature of 49 500 $\rmn{K}$, where the {\it rss\/} value was at a minimum. And, the error in 
the temperature of the hot component was estimated as 4500 K from 
the examination of $T_{1}$ -- {\it rss\/} variation. The observed and computed light curves are given in Fig.4 and the parameters are listed in Table 2.   
\begin{figure}
\center
 \includegraphics[width=120mm]{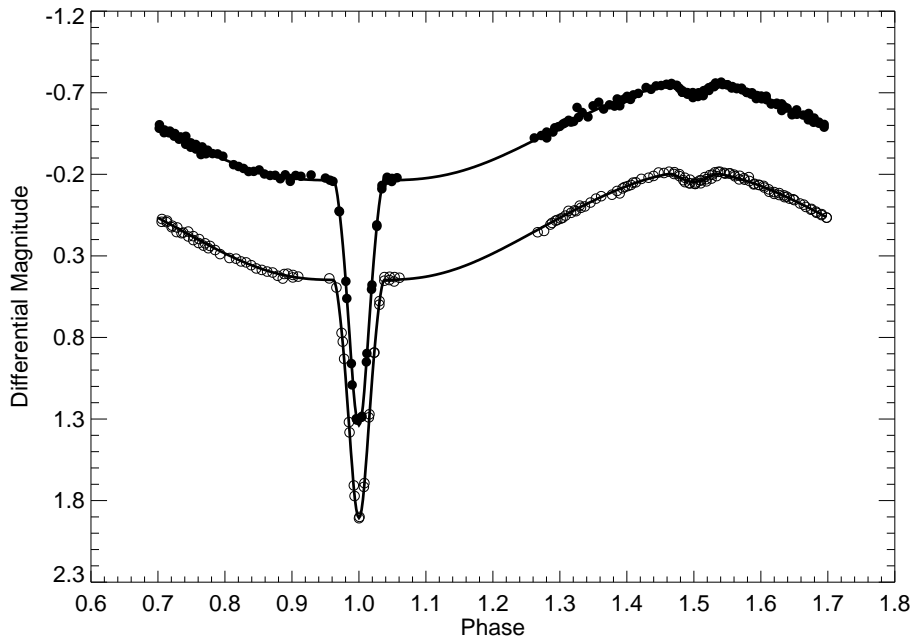}
 \caption{{\it V--\/} and {\it R--\/}passband light curves of V477 Lyr with computed fits.}
\end{figure}

	The linear limb-darkening coefficients derived from the solution exhibit a noteworthy discrepancy from those published ones that mentioned above. The cool component acts like a very 
hot star due to the reflection effect caused by the subdwarf primary. The heating on the hemisphere of the secondary star that faces to hot primary is such vigorous that the irradiation 
affects the limb-darkening over the stellar surface. Considering the grazing nature of the secondary eclipse, \citet{b11} made an interpretation on the effectiveness of the reflection 
effect on V477 Lyr and suggested that the horizontal energy transport in the atmosphere of the secondary component might be the reason for the brighter limb.
	
	To determine the temperature of the subdwarf companion, \citet{b43} computed the Zanstra temperatures using the H and He\,{\sc ii} nebular lines and constrained the primary 
component's temperature in between 32 000 and 60 000 $\rmn{K}$. In a most recent study, PB94 presented a new Zanstra analysis based on the H$\beta$ and He\,{\sc ii} nebular lines and 
estimated the temperatures as 48 000 and 82 000 $\rmn{K}$, respectively. They also employed the R--technique of \citet{b69} and derived a temperature of 63 000 $\pm$12 000 $\rmn{K}$ for 
the primary star and considering the temperature range found from several methods they decided to adopt a temperature of 60 000$\pm$10 000 $\rmn{K}$ in the light curve analysis. However, 
the simultaneous two--colour photometry gives us an opportunity to derive more accurate and approximate values for the temperatures of the binary components, which we have found as 49 
500 $\rmn{K}$ (fixed) for the primary and 3874$\pm$120 $\rmn{K}$ for the secondary star. The uncertainty of 120 $\rmn{K}$ is the formal 1$\sigma$ error from the solution. The corrected 
error can be estimated as 350 $\rmn{K}$ based on the uncertainty of 4500 $\rmn{K}$ in the effective temperature of the primary. The computed astrophysical parameters are given in Table 
3. In fact, the mass of the secondary component corresponds to an M type star (as in the case of UU Sge) and a lower temperature is required for a main--sequence star at the same mass. 
One of the reasons why we could not reach such an expected temperature is the limb--brightening phenomenon, which takes place on the overheated hemisphere of the cool components.

\begin{table*}
\centering
\caption{Astrophysical data for UU Sge and V477 Lyr.}
 \begin{tabular}{@{}lcc@{}}
   \hline
   \hline
Parameter&UU Sge&V477 Lyr\\
\hline
$M_{1}$	(M$_{\odot}$)&	0.628	$\pm$	0.053	&	0.508	$\pm$	0.046\\
$M_{2}$	(M$_{\odot}$)&	0.288	$\pm$	0.031	&	0.145	$\pm$	0.021\\
$R_{1}$	(R$_{\odot}$)&	0.349	$\pm$	0.013	&	0.172	$\pm$	0.007\\
$R_{2}$	(R$_{\odot}$)&	0.558	$\pm$	0.021	&	0.462	$\pm$	0.019\\
$T_{1}$ ($\rmn{K}$) & 78 000 $\pm$ 3000 & 49 500 $\pm$ 4500\\
$T_{2}$ ($\rmn{K}$) & 6136 $\pm$ 240 & 3874 $\pm$ 350 \\
$\log g_{1}$ ({\it cgs\/})	&	5.151	$\pm$	0.050	&	5.672	$\pm$	0.053\\
$\log g_{2}$ ({\it cgs\/})	&	4.404	$\pm$	0.058	&	4.269	$\pm$	0.073\\

  \hline
 \end{tabular}
 \end{table*}

\section{The Reflection Effect and the Enhanced Effective Temperatures of the Secondary Components}

	The hot subdwarf components of UU Sge and V477 Lyr, which are also responsible for the excitation of the nebulae, overheat the hemispheres facing to the substellar points of the cool 
components and the reflection effect reveals itself on the light curves' maxima as a sine--like distortion with large amplitudes (Figs. 1 and 2). The irradiation on the secondary 
components significantly increases the temperature around the substellar points. To estimate the reflection rates we applied Fourier analysis to the out-of-eclipse variation of the light 
curves and the following truncated Fourier series was used to compute the Fourier coefficients,
\begin{equation}
  l = A_{0} + A_{1}cos\theta + A_{2}cos2\theta + B_{1}sin\theta + B_{2}sin2\theta. 
\end{equation}
In this equation, {\it l} is the normalized intensity, ${\theta}$ is the phase in degrees, $A_{0}$ is the normalization level for the intensity, $A_{1}$ and $A_{2}$ are the contribution 
to the reflection effect. $B_{1}$ and $B_{2}$ are the coefficients represent the other perturbations on the light curve. The normalized Fourier coefficients of UU Sge and V477 Lyr for 
both passbands are listed in Table 4. The $B$ coefficients have significantly smaller values than other coefficients, which indicate that other perturbations have no remarkable influence 
on the light curves other than reflection effect.

\begin{table*}
\centering
\caption{Normalized Fourier coefficients ({\it l}=1 at ${\theta}$=90$\degr$).}
 \begin{tabular}{@{}lccccccc@{}}
   \hline
   \hline
Targets&passbands&$A_{0}$&$A_{1}$&$A_{2}$&$B_{1}$&$B_{2}$\\
\hline
UU Sge&{\it V\/}&   1.0265&	-0.1907&0.0312&	0.0062&	-0.0002\\
& & \,\,\,\, $\pm$15& \,\,\,\,\, $\pm$22& \,\,\,\, $\pm$22& \,\,\,\, $\pm$15& \,\,\,\,\, $\pm$17\\
	&	{\it R\/}	&	1.0251	&		-0.2147	&	0.0356	&		0.0110	&	0.0016	\\
& & \,\,\,\, $\pm$13	& \,\,\,\,\, $\pm$19	&	\,\,\,\, $\pm$18	& \,\,\,\, $\pm$12	&	\,\,\,\, $\pm$14\\
\hline
V477 Lyr	&	{\it V\/}	&	1.0477	&		-0.2912	&		0.0540	&		0.0053	&		0.0027		\\
& & \,\,\,\, $\pm$22	& \,\,\,\, $\pm$33	& \,\,\,\,\, $\pm$25	& \,\,\,\, $\pm$30	& \,\,\,\, $\pm$29\\
	&	{\it R\/}	&	1.0630	&		-0.3338	&		0.0617	&	-0.0011	&	0.0015		\\  
& & \,\,\,\, $\pm$14	& \,\,\,\,\, $\pm$20	& \,\,\,\, $\pm$17	&	\,\,\,\,\, $\pm$19	&	\,\,\,\, $\pm$18\\
  \hline
 \end{tabular}
 \end{table*}

	We have computed the amplitudes of the sine--like distortions caused by the reflection effect by using the Fourier coefficients of related systems. The amplitudes for UU Sge were 
found to be 0.396 mag and 0.446 mag in {\it V--\/} and {\it R--\/}passband, respectively. The amplitude found in {\it V--\/}passband light curve seems in very good agreement with that of 
derived by PB93. In the case of V477 Lyr, the amplitudes in {\it V--\/} and {\it R--\/}passband were estimated to be 0.588 mag and 0.665 mag, in fair agreement with that of obtained by 
PB94.

We have tried to estimate the maximum temperature values, $T_{2h}$, around the substellar points of the secondary components by employing the following equation; 
\begin{equation}
  T_{2h}^{4} = \frac{1}{R_{2}^{2}\phi^2}\Big[10^{-0.4\Delta{m}}(R_{1}^{2}T_{1}^{4} + R_{2}^{2}T_{2}^{4}) - R_{1}^{2}T_{1}^{4}\Big]. 
\end{equation}

Where $R_{1}$, $R_{2}$ are the radii and $T_{1}$, $T_{2}$ are the effective temperatures of the primary and the secondary components, respectively; $\Delta{m}$ is the amplitude of the 
brightness variation from minimum ($\theta$=0$\degr$, mid-primary) to maximum ($\theta$=180$\degr$, mid-secondary) of the sine--like distortion, and $\phi$ is the approximate central 
angle viewed from the centre of the secondary, which corresponds to the irradiated area over the surface of the secondary star (see \citealt{b89}). We have determined the {\it V--\/} and 
{\it R--\/}passband substellar--point temperatures of the cool components of UU Sge to be around 30 680 $\rmn{K}$ and 31 800 $\rmn{K}$, and of V477 Lyr to be around 16 860 and 17 550 
$\rmn{K}$, respectively. These estimates agree well with those of computed by the equation described in \citet{b89}.

\section{Oversized Radii of the Secondary Components and Evolutionary Status of the Systems}
	
	Radii of the secondary components of UU Sge and V477 Lyr have been found to be larger than those expected for main--sequence stars at the same masses. On the other hand, our targets 
are not the only examples for such systems that contain oversized secondaries. In this section, we discuss the possible reasons on the oversized nature of the late-type secondary 
components of our targets along with some other selected PNNi and post--CE systems (listed in Table 5). Although the mass of the secondary component of V471 Tau is almost three times the 
masses of other secondaries studied here, we have especially included this well--studied system for comparison. In a study made by \citet{b58}, two possible reasons for the expanded 
radius of the K--type secondary of V471 Tau were accentuated: (1) because of the possible CE phase, which V471 Tau might have recently emerged from, the secondary star may be still out 
of thermal equilibrium or (2) the starspots covering the surface of the star prevent the convective energy from being transported, which as a result increase the radius of the star.

The CE evolution and its effect on the low--mass main--sequence secondary components are the phenomena that have been poorly understood both theoretically and observationally. If a close 
binary system, consisting of a WD primary and a red dwarf (RD) secondary, have once undergone a CE phase, the mass and the radius of the secondary component might have been changed due 
to mass transfer during the formation of the CE. Although there remain questions on whether the mass of the secondary component alters dramatically or does not change in a substantial 
manner, there are observational evidences which show the significant changes in the radius of the component, such as the systems given in this paper. 

There are several theoretical studies that have focused upon the response of the secondary component to the accretion process during the CE evolution. \citet{b62} made some evolutionary 
calculations for a 0.2 M$_{\odot}$ fully convective star which accretes matter at several rates and investigated the outcome of the accretion. They found that in the models with 
accretion rates around 10$^{-4}$--10$^{-5}$ M$_{\odot}$ yr$^{-1}$, the incoming material accumulates faster than the contraction rate, and consequently the outer radius increases. In 
their models the final value of the accumulated matter reaches to 2.5$\times$10$^{-3}$ M$_{\odot}$, which basically indicates that the mass of the fully convective star remains almost 
unchanged. They also concluded that the results should be valid for all fully convective stars ($\la$0.3 M$_{\odot}$) as well as for the stars with deep convective envelopes (0.3 
M$_{\odot}$$\la$$M$$\la$0.5 M$_{\odot}$). In a similar study carried out by \citet{b66}, the effect of rapid mass accretion onto low--mass main--sequence stars (0.3 M$_{\odot}$--0.5 
M$_{\odot}$) was investigated. They, too, came to a similar conclusion and indicated that the stars with deep convective envelopes experiencing rapid mass accretion significantly 
increase their radii. Another study published by \citet{b34} also probed the response of a 1.25 M$_{\odot}$ main--sequence star to the conditions within the CE. At a range of relatively 
high accretion rates (10$^{-5}$ to 10$^{-1}$ M$_{\odot}$ yr$^{-1}$), they found that the radius of the secondary component can expand ten times larger than its main--sequence radius upon 
mass accretion of less than 0.1 M$_{\odot}$. After filling its Roche lobe, the secondary loses back almost all the mass it has accreted and only a small amount of mass ($\la$0.01 
M$_{\odot}$) remains accumulated after emergence from the envelope. As a result, the surface layers of the secondary depart from thermal equilibrium and enter in a thermally relaxation 
stage with a radius larger than its initial main--sequence radius. On the other hand, the results of some three--dimensional hydrodynamical simulations (e.g. \citealt{b65}) reveal that 
the evolution of the CE phase takes place in short dynamic time scales and the secondary star which moves rapidly (supersonic) in the CE leaves this stage almost without mass accretion. 
However, the cases we will discuss in the following paragraphs involve the possible small amount of mass accretion by the cool companion.   

The second reason which has been considered as a process to increase the radii of the low-mass secondary components is the effectiveness of the magnetic activity. There are several 
studies on low--mass main--sequence stars, which compare the physical properties of magnetically active and inactive stars. \citet{b56} showed that, among the stars at the same masses, 
active stars are cooler and have larger radii than inactive stars. In a most recent study \citet{b55} concluded that all active low--mass stars (including low--mass eclipsing binaries) 
are cooler than inactive stars of similar luminosity and hence have larger radii.

Recently, \citet{b23} have discussed the importance of the irradiation effect in binary PNNi and suggested that the heating which takes effect on the `day side' of the secondary stars 
might cause a puffing up and expand the surface layers of the secondaries.  

By taking into consideration two of the reasons summarized above, we have carried out some calculations to study the evolutionary status of 
some selected post--CE systems and interpreted which reason may be responsible for the cool components to have oversized radii: incomplete thermal relaxation or presence of magnetically 
active regions.
 
{\it KV Vel} : The binary nucleus of the PN DS 1. The sdO star was discovered by \citet{b26} and double--lined spectroscopic variations with a period of 8.571 hour were found by 
\citet{b27}. Later, the renewed results for the system parameters were published by \citet{b33}.

{\it BE UMa} : It is an eclipsing binary nucleus. Although the light variability of the system was first noted by \citet{b46} and an eclipse was first observed by \citet{b4}, the nebula 
surrounds the binary nucleus was discovered by \citet{b52}. The system parameters listed in Table 5 and 6 have been adopted from \citet{b29}.

{\it HS 1136+6646} : Discovered as a spectroscopic binary by \citet{b32} and more recently \citet{b70} have indicated that the double--lined spectroscopic system is a newly formed 
post-CE binary. 

{\it PG 1026+002} : The system, which consists of a WD primary and a RD secondary, was discovered during the Palomar Green Survey (\citealt{b30}).
\citet{b64} reported that the system is a single--lined spectroscopic binary and the secondary component's radius is compatible with its mass. However, by using the {\it R--\/}passband 
light curve of the system \citet{b15} found the radius of the secondary as being larger than expected from the theoretical models.

{\it RR Cae} : An eclipsing WD--M dwarf binary system with a period of 7.3 hour. The eclipses were first observed by \citet{b45}. \citet{b14} analysed the radial velocity measurements 
together with the photometric data of \citet{b13} and found that the radius of the M dwarf is larger as compared with a main--sequence star at the same mass. However, in a more recent 
study, \citet{b54} presented new photometry and double--lined spectroscopy of the system and concluded that, contrary to \citet{b14}, the radius of the secondary is consistent with its 
mass. By taking into account the discrepancy between these two results we have performed two separate calculations.

{\it RE 1016-053} : Post-CE binary consists of a WD primary and a RD secondary. Two different values for the secondary star's radius were given by \citet{b82} and \citet{b76} depending 
on two different spectral analysis and the radius of the secondary component was found as slightly oversized. Both radii have been considered here and the results are listed in Table 5.

{\it V471 Tau} :  Is a well-known post-CE binary system (WD primary + RD secondary) and a member of the Hyades star cluster. The system has a period of 12.5 hour and being observed since 
the discovery of its eclipses by \citet{b57}. \citet{b37} showed that the RD secondary is an active star. Possible reasons for the expanded radius of the secondary component were 
discussed by \citet{b58} and they attributed the expansion to the magnetic activity which originates from the starspots partially covering the surface of the star.            

\begin{figure}
\center
 \includegraphics[width=120mm]{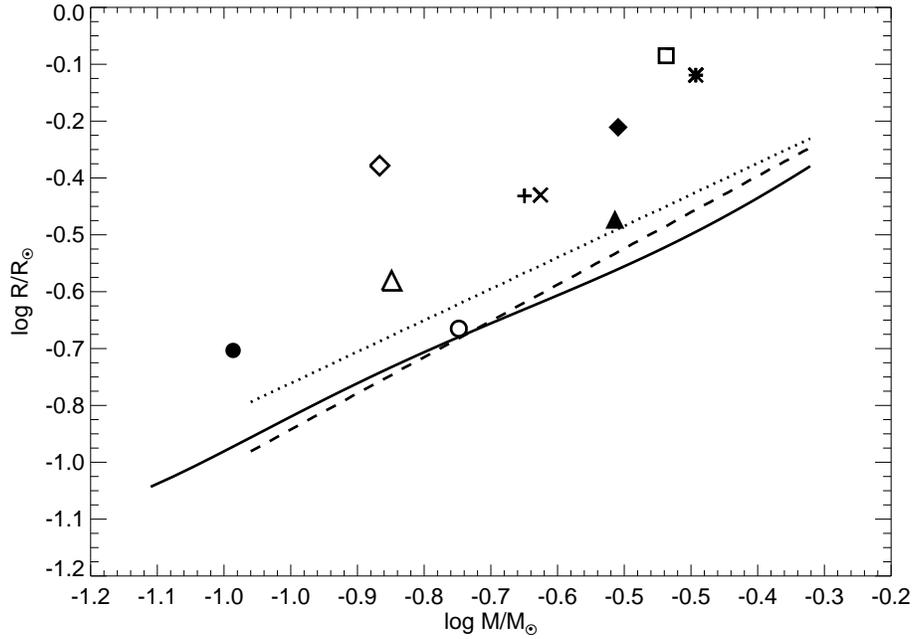}
 \caption{Mass--radius relationships for the low mass main--sequence stars adopted from different studies. The fits are taken from: solid--line: CB97; dotted--line: \citet{b17}; 
dashed--line: \citet{b50}. Radii of the secondary components are plotted as; UU Sge: filled diamond; V477 Lyr: open diamond; KV Vel: cross; BE UMa: asterisk; HS 1136+6646: open square; 
PG 1026+002: plus; RR Cae (\citealt{b14}): filled circle; RR Cae (\citealt{b54}): open circle; RE 1016-053 (\citealt{b82}): filled triangle; RE 1016-053 (\citealt{b76}): open triangle.}
\end{figure}
\begin{figure}
\center
 \includegraphics[width=120mm]{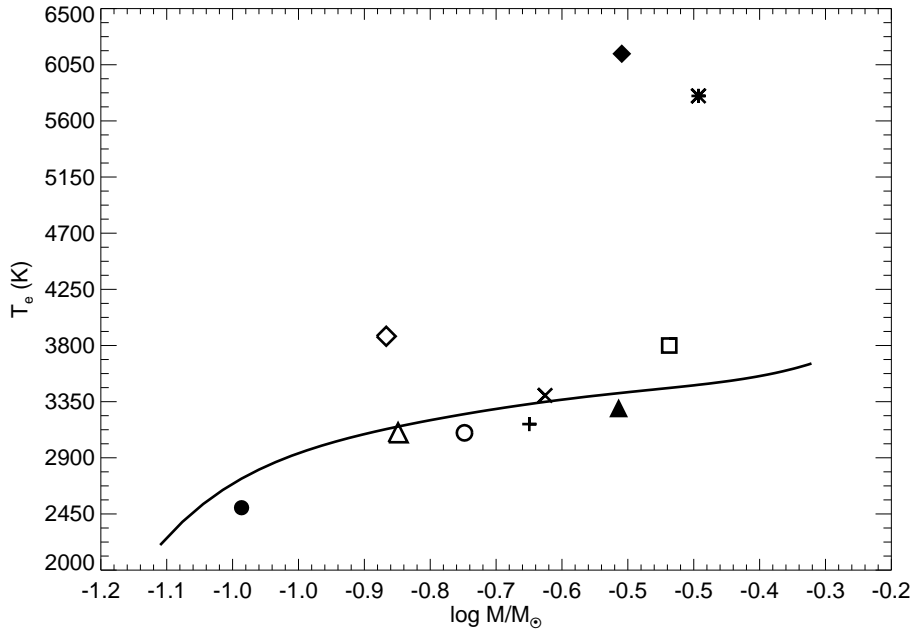}
 \caption{Mass--effective temperature relationship for the low mass main--sequence stars (CB97). The symbols are the same as in Figure 5.}
\end{figure}
\begin{figure}
\center
 \includegraphics[width=120mm]{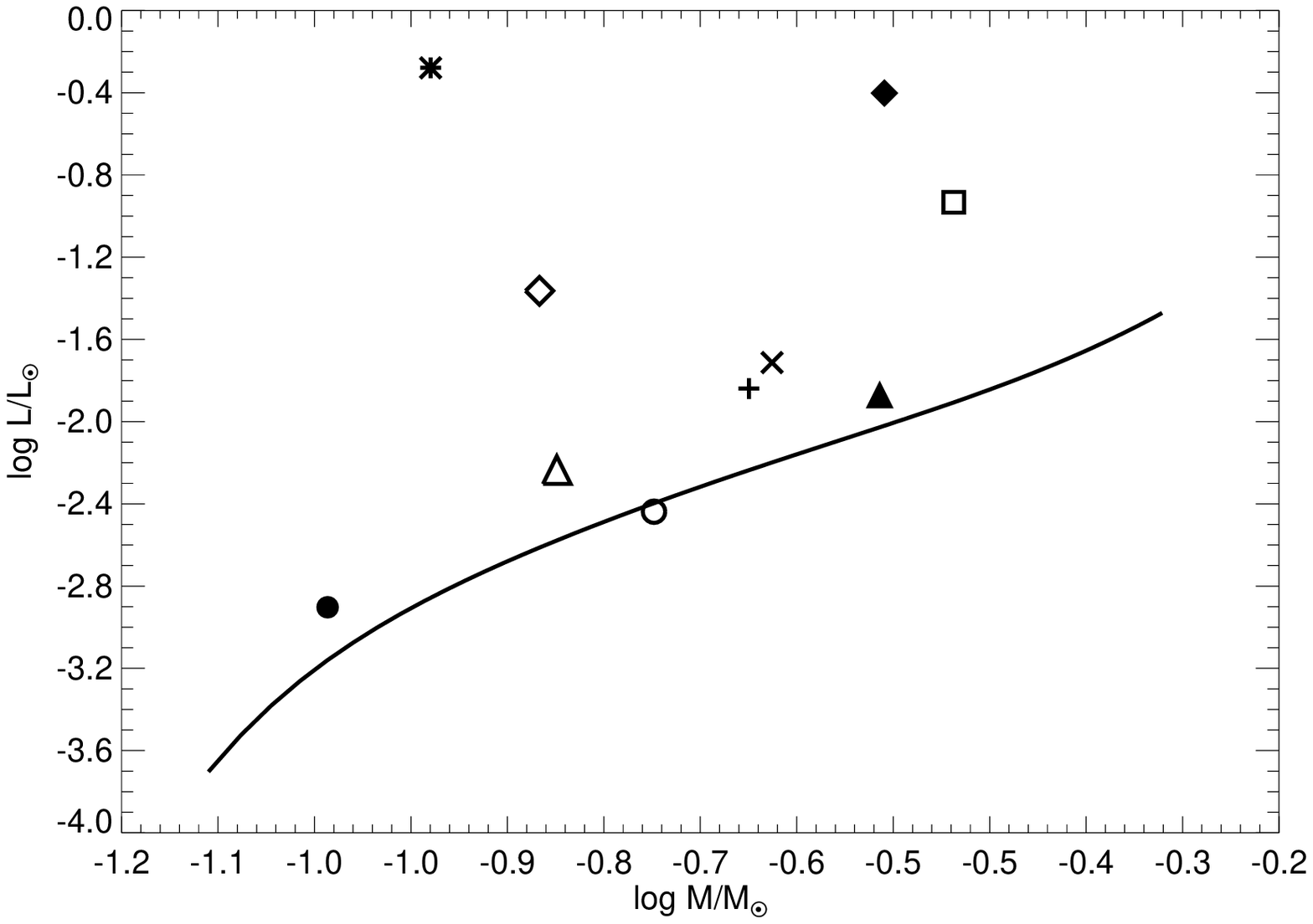}
 \caption{Mass--luminosity relationship for the low mass main--sequence stars. The symbols are the same as in Figure 5.}
\end{figure}

We have used the models (age of 1 Gyr, [M/H]=0.0) provided by \citet{b18} to find the ``$expected$" radii, temperatures and luminosities of the secondary components. The models have been 
fitted by sixth-degree polynomials and these equations have been used to calculate the ``$expected$" radii ($R_{2,teo}$) given in Table 5 (along with ``$observed$" ones). The 
mass--radius relationship of CB97 is plotted in Figure 5 and comparison is made with the models of \citet{b50} and \citet{b17}. The ratio of the ``$observed$" and ``$expected$" radii are 
also given in Table 5. As it is seen from these ratios, most of the secondary components, including UU Sge and V477 Lyr, have relatively larger radii (2 or higher) than the radii of 
their main--sequence equivalents.

The locations of the secondary companions are illustrated in the mass--temperature ($M$--$T_{\rmn{e}}$) and mass--luminosity ($M$--$L$) planes in Figs. 6 and 7. Since the effective 
temperatures could not be determined for the secondary stars of HS 1136+6646 and RE 1016-053, the suggested spectral types taken from the referred papers are given in Table 5. In order 
to plot these stars in the $M$--$T_{\rmn{e}}$ and $M$--$L$ planes we adopted the effective temperatures from the tables given by \citet{b51}. The temperatures of 3800 $\rmn{K}$ for K7 
classification of HS 1136+6646, 3300 $\rmn{K}$ and 3100 $\rmn{K}$ for M3 and M5 classifications of RE 1016-053 were used in our calculations.       

\begin{table*}
\caption{The observed and theoretical properties of the secondary components of the selected post-CE systems. The description of the properties are made in this section.}
 \begin{tabular}{@{}l|c|c|c|c|c|c|c|c|c|c@{}}
   \hline
   \hline

Object	&	$M_{2}$	&	$R_{\rmn{2,obs}}$	&	$T_{2}$ & $T_{\rmn{2,teo}}$ &	$R_{\rmn{2,teo}}$ &	$R_{\rmn{2,obs}}/R_{\rmn{{2,teo}}}$	&	$t_{\rmn{th,0.05}}$ 	&	
$t_{\rmn{th,0.01}}$	&	$t_{\rmn{th,0.005}}$& Ref. \\

&	$_{(M_{\odot})}$& $_{(R_{\odot})}$	& $_{({\rmn{K}})}$	& $_{({\rmn{K}})}$ &	$_{(R_{\odot})}$&  &$_{(yr)}$& $_{(yr)}$& $_{(yr)}$ & \\
 \hline
UU Sge	&	0.288	&	0.558	&	6138 & 3420&	0.280	&	1.989	&	1.64$\times$10$^{8}$	&	3.27$\times$10$^{7}$	&	1.64$\times$10$^{7}$ & this work\\
V477 Lyr	&	0.145	&	0.463	&	3874 & 3130 &	0.170	&	2.727	& 9.32$\times$10$^{8}$	&	1.86$\times$10$^{8}$	&	9.32$\times$10$^{7}$	& this work\\
KV Vel	&	0.230	&	0.402	&	3400 & 3340 &	0.238	&	1.688	&	1.62$\times$10$^{8}$	&	3.24$\times$10$^{7}$	&	1.62$\times$10$^{7}$ & 1\\
BE UMA	&	0.360	&	0.720	&	5800	& 3480 &	0.336	&	2.142	&	1.28$\times$10$^{8}$	&	2.56$\times$10$^{7}$	&	1.28$\times$10$^{7}$ & 2\\
HS 1136+6646	&	0.330	&	0.790	&	K7	& 3460	& 0.312	&	2.530	&	2.01$\times$10$^{8}$	&	4.03$\times$10$^{7}$	&	2.01$\times$10$^{7}$ & 3\\
PG1026+002	&	0.220	&	0.400 &	3170	& 3330 &	0.231	&	1.733	&	1.85$\times$10$^{8}$	&	3.71$\times$10$^{7}$	&	1.85$\times$10$^{7}$ & 4,5\\
RR Cae	&	0.095	&	0.189	&	2500	& 2730 &	0.118	&	1.598	&	1.07$\times$10$^{9}$	&	2.15$\times$10$^{8}$	&	1.07$\times$10$^{8}$ & 6\\
RR Cae	&	0.182	&	0.210	&	3100	& 3250 &	0.202	&	1.042	&	1.45$\times$10$^{7}$	&	2.90$\times$10$^{6}$	&	1.45$\times$10$^{6}$ & 7\\
RE 1016-053	&	0.285	&	0.357	&	M3	& 3420 &	0.278	&	1.283	&	4.76$\times$10$^{7}$	&	9.53$\times$10$^{6}$	&	4.76$\times$10$^{6}$ & 8\\
RE 1016-053	&	0.150	&	0.265	&	M5	& 3150 &	0.174	&	1.521	&	2.62$\times$10$^{8}$	&	5.24$\times$10$^{7}$	&	2.62$\times$10$^{7}$ & 9\\
V471 Tau	&	0.930	&	0.960	&	5040	& 5370 &	0.832	&	1.153	&	5.15$\times$10$^{5}$	&	1.03$\times$10$^{5}$	&	5.15$\times$10$^{4}$ & 10\\
  \hline
 \end{tabular}
\begin{list}{}{}
\item[{References:}] The references are given for the observed properties of the secondary components. (1) \citet{b33}, (2) \citet{b29}, (3) \citet{b70}, (4) \citet{b15}, (5) 
\citet{b64}, (6) \citet{b14}, (7) \citet{b54}, (8) \citet{b82}, (9) \citet{b76}, (10) \citet{b58}.   
 \end{list}
 

 
\caption{The properties of the primary components of the selected post-CE systems.}
 \begin{tabular}{@{}l|c|c|c|c|c|c|c|c@{}}
   \hline
   \hline

Object	&	$M_{1}$	&	$T_{1}$ 	&	$\log g_{1}$ 	 &	 $Age$ &  Ref.\\
 & $_{(M_{\odot})}$ & $_{({\rmn{K}})}$	& $_{({\it cgs\/})}$	& $_{(yr)}$ & \\
 \hline										
UU Sge	&	0.628	&	78 000	&	5.151	 &	3.0$\times$10$^{4}$ & 1\\	
V477 Lyr&	0.508	&	49 500	&	5.672	 &	3.0$\times$10$^{4}$ & 1\\		
KV Vel	&	0.63	&	77 000	&	5.85	 &	3.0$\times$10$^{4}$ & 1 \\		
BE UMA	&	0.70	&	105 000	&	6.5	 &	3.0$\times$10$^{4}$ & 1 \\		
HS 1136+6646&	0.63	&	70 000	&	7.75	 &	9.0$\times$10$^{5}$ & 2\\		
PG1026+002&	0.65	&	17 600	&	8.1	 &	1.4$\times$10$^{8}$ & 2\\		
RR Cae&	0.467	&	7005	&	7.7	 &	1.9$\times$10$^{9}$ & 3\\		
RR Cae&	0.44	&	7540	&	7.7	 &	1.9$\times$10$^{9}$ & 3\\		
RE 1016-053&	0.57	&	55 800	&	7.81	 &	1.8$\times$10$^{6}$ & 2\\		
RE 1016-053&	0.59	&	55 000	&	7.84	 &	1.8$\times$10$^{6}$ & 2\\		
V471 Tau	&	0.84	&	34 500	&	8.31	 &	1.2$\times$10$^{7}$ & 2\\		
 
 \hline
 \end{tabular}
\begin{list}{}{}
\item[{References:}]  The references given here are for the models which have been used to compute the ``age" of the systems (for the other properties given in this table please see the 
references given in Table 5). (1) \citet{b68}, (2) AB98, (3) BA98.
 \end{list}
 
 \end{table*}

In order to have an idea on the thermal relaxation time--scale ($t_{\rmn{th}}$) of the envelope (accumulated matter) which was
formed during the mass transfer from the primary to the secondary, we have used $t_{\rmn{th}}$ equation given by \citet{b62}:

\begin{equation}
  t_{\rmn{th}} \simeq GMM_{\rmn{env}}\Delta{R} / R^{2}L. 
\end{equation} 

Where $M$ and $R$ is the secondary's mass and ``$expected$" radius, $M_{\rmn{env}}$ is the mass of the envelope due to mass accretion, $\Delta{R}$ is the width of
the envelope and $L$ is the bolometric luminosity of the secondary star.

Following the theoretical studies mentioned in earlier paragraphs we have assumed three $M_{\rmn{env}}$ values which may have been accumulated 
by the secondary star: 0.05, 0.01 and 0.005 M$_{\odot}$. Then we have calculated the $t_{\rmn{th}}$ for each assumption. The results are given in Table 5 as $t_{\rmn{th,0.05}}$, 
$t_{\rmn{th,0.01}}$ and $t_{\rmn{th,0.005}}$.           

Since it is very important to know the critical accretion rates, we have also calculated the Eddington luminosities of the secondary stars. We have found the critical accretion 
rates to be of the order of a few 10$^{-4}$ M$_{\odot}$ yr$^{-1}$. Therefore, it is acceptable to consider a mass accretion onto secondary components as rapid as a few 10$^{-5}$ 
M$_{\odot}$ yr$^{-1}$, which is an adequate accretion rate leads to increase in the radii of the cool components as previously discussed by \citet{b62}.

For the binary nuclei, UU Sge, V477 Lyr, KV Vel and BE UMa, the presence of the PN warrants us in estimating the ages (the time since the emergence 
from the CE) of the systems as a few 10$^{4}$ yr (30 000 -- 40 000 yr, \citealt{b68}, \citealt{b71}). So, we have simply adopted the ages as 3$\times$10$^{4}$ yr for all the nuclei  
(Table 6). However, we have used the evolutionary tracks computed by \citet{b7}, and showed the possible evolutionary states of the sdO componets of UU Sge and V477 Lyr in Fig. 8. 
Although these models were computed for PNe with single nuclei, the loci of the sdO components seem in fairly well agreement (within the uncertainties given for the masses of sdO 
components) with the models have remnant masses of 0.625 and 0.524 M$_{\odot}$, which
evolved from the initial ZAMS masses of 3 and 1 M$_{\odot}$, respectively. In these models, the remnants have CO cores and the nuclear energy is mainly provided by the He shell burning. 
Besides all these approaches, we are aware of the fact that our systems are not PNe with ``single" nuclei, and binarity may have influenced the evolution of the components through the 
interaction between the components and formation of the CE (more details on the evolution of PNe and CE can be found in the referred papers), as much as the different mass--loss rates 
the systems might have encountered.  

\begin{figure}
\center
 \includegraphics[width=120mm]{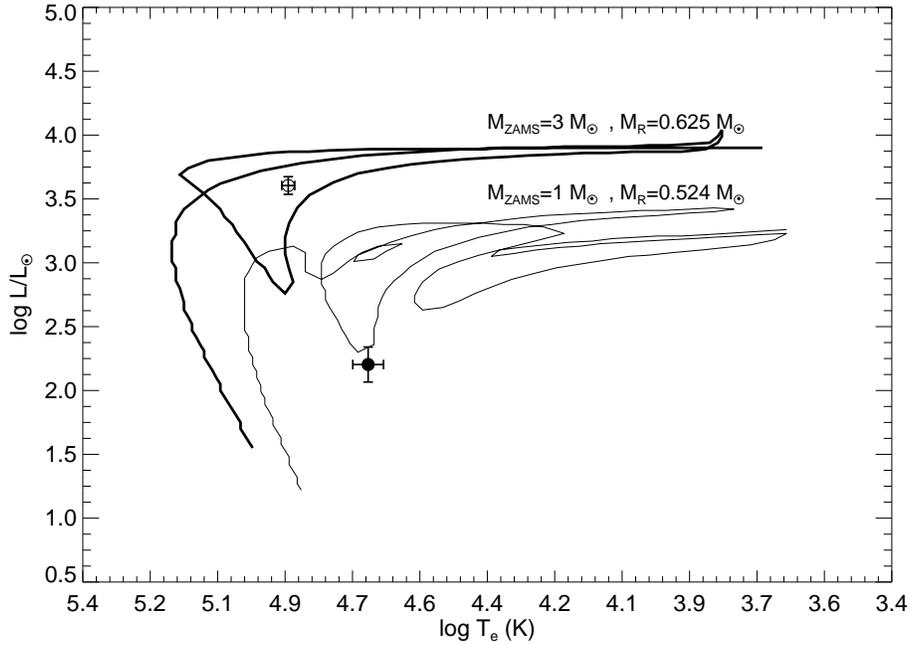}
 \caption{The evolutionary status of UU Sge and V477 Lyr. The models are taken from \citet{b7}. Open and filled circles represent the primary component of UU Sge and V477 Lyr, 
respectively.}
\end{figure}

For the other systems studied here we have used the white dwarf evolutionary models from \citet{b3} and \citet{b6} to calculate the approximate cooling ages, $t_{\rmn{cool}}$.
In AB98, the evolutionary models were computed for the carbon--oxygen WD masses within the range of 0.5--1.2 M$_{\odot}$ and surface hydrogen layer masses ranging between 10$^{-13}$ and 
10$^{-4}$ $M_{\rmn{H}}$/$M$, where {\it M} is the mass of the modeled WD and $M_{\rmn{H}}$ is the mass of the 
H envelope. BA98 developed the models for the evolution of helium WDs with masses ranging from 0.15 to 0.5 M$_{\odot}$, and having outer H envelope 
masses within the range of 10$^{-8}$ $\leq$$M_{\rmn{H}}$/$M$$\leq$ 4$\times$10$^{-3}$. And both WD models were improved for the metallicities of $Z$=0.0 and $Z$=0.001.

For the primary masses higher and lower than 0.5 M$_{\odot}$, we have used the tables given by AB98 and BA98, respectively. 
Various $M_{\rmn{H}}$/$M$ and both $Z$ values have been taken into account and the models have been examined in $T_{\rmn{e}}$--$\log {g}$ plane. The most
applicable models have been selected and interpolated for the related primary masses in order to estimate the cooling ages of the WDs. We have found that the models for V471 Tau and RR 
Cae having the surface hydrogen layer mass of 10$^{-4}$ $M_{\rmn{H}}$/$M$ give the best agreement in $T_{\rmn{e}}$--$\log {g}$ plane for the metallicities of $Z$=0.0 and $Z$=0.001, 
respectively. However, other systems seem consistent with the models have surface hydrogen layer mass of 10$^{-6}$ $M_{\rmn{H}}$/$M$ and metallicity of $Z$=0.001. Resultant 
$t_{\rmn{cool}}$ values are listed in Table 6.     
    
Since our goal is to search for the more plausible reason on the oversized nature of the secondaries, we have compared $t_{\rmn{th}}$ with the ages of the systems as summarized below:       	

{\it UU Sge}, {\it V477 Lyr}, {\it KV Vel} and {\it BE UMa} : The thermal relaxation time--scales calculated for each of the envelope masses are  relatively longer than the ages of the 
systems, which indicate that the surface of the secondary components have not yet reached the thermal equilibrium. The discrepancy between the ``$observed$" and the ``$expected$" radii 
of the secondaries can be seen in Fig. 5. The difference of the effective temperatures from the theoretical ones is also notable (Fig. 6).                

{\it HS 1136+6646} : The comparison of $t_{\rmn{th}}$ and $t_{\rmn{cool}}$ of the system shows that surface layers of the system is still in a thermally relaxation stage which causes the 
expansion in the radius of the secondary. Also the adopted effective temperature which corresponds to a K7 star is located above the theoretical fit and appears to be in agreement with 
this approach.

{\it PG 1026+002} : When the computed $t_{\rmn{th}}$ values are compared with $t_{\rmn{cool}}$ of the system, depending on the envelope mass, the secondary component seems to be either 
at the end of its thermally relaxation stage or just passed it.
However, the low effective temperature of the secondary component may be taken as support for the idea that the reason for the expanded radius is the presence of the possible 
magnetically active regions on the secondary star.    

{\it RR Cae} : For the parameters given by \citet{b14}, the thermal relaxation time--scales for the secondary component of RR Cae are shorter than the cooling age of the system. The 
inconsistency both in radius and effective temperature with the theoretical expectations indicates the presence of the possible stellar activity. On the other hand, the posibility of the 
higher amount of matter (at least 0.1 M$_{\odot}$) being accreted by the secondary leaves a question mark that the cool component might be still out of thermal equilibrium. \citet{b54} 
found the radius consistent with the theoretical expectations as seen in Fig. 5. But there seems to be a disagreement between effective temperature and its expected counterpart, which 
also supports the assumption of the presence of active regions. Because the mass of the secondary component is twice as much of the mass found by \citet{b14}, $t_{\rmn{th}}$ values are 
relatively well below the cooling age of the system.

{\it RE 1016-053} : There is an interesting result for the secondary star of RE 1016-053. The radius of the cool component (taken from both \citealt{b82} and \citealt{b76}) has larger 
values than expected according to the mass--radius relationship of CB97 and the comparison of the thermal relaxation time--scales with the cooling age of the system shows that the 
secondary star still may have thermal instability on its surface layers. But the adopted effective temperatures seem to be a little lower than expected, which could tell us the 
possibility of having active regions on the surface of the secondary if we would have lower $t_{\rmn{th}}$ values than $t_{\rmn{cool}}$. However, our calculations indicate a secondary 
star which is far from being thermally relaxed that makes us to expect higher effective temperatures. Since the effective temperatures have been determined from the spectral types 
estimated by the authors (\citealt{b82} and \citealt{b76}) and they suggest these spectral types relying on the masses of the secondary star, we believe this is due to probable 
underestimation of either spectral types or the masses.

{\it V471 Tau} : Our findings on this well-known post-CE system are listed in Table 5 and 6, but since the mass of the secondary component is higher than the rest of the secondaries 
studied here, it is not plotted in Figs. 5, 6 and 7. It is obvious that the secondary component passed its thermally relaxation stage relatively long ago. Nonetheless, its radius seems a 
little larger and the effective temperature seems a little lower. These results are in fairly well agreement with those presented by \citet{b58}.     
  
\section{The Efficiency Parameter : ``$\alpha$$_{\rmn{CE}}$"}

The CE evolution is one of the most crucial part in the evolution of binary systems. The presence of close binaries, for example, consisting
of a WD primary and a RD secondary, indicates that the wide precursor system must have undergone a significant amount of angular momentum
loss which causes the orbital shrinkage. 

At some stage of the evolution, more massive primary component evolves to AGB (or RGB, depending on the initial configuration of the system) 
and fills its Roche lobe which leads to mass transfer towards the secondary component. If the mass transfer is unstable the secondary star can not accommodate the accreted matter at its 
surface and both components are engulfed by the primary's envelope which is later on being called as CE. Since the CE does not corotate with the orbiting binary core, the frictional 
forces arise due to velocity difference. This frictional 
interaction produces gravitational drag that causes components to spiral-in together and reduces the binary separation. The change in orbital separation results in the change in orbital 
energy and a fraction of the difference in orbital energy is used to eject the envelope from the system by depositing it into CE. This fraction is designated as the ``{\it efficency}" 
parameter and denoted by ``$\alpha$$_{\rmn{CE}}$" (e.g. \citealt{b73}, \citealt{b41}, and references therein)

\begin{equation}
  \alpha_{\rmn{CE}} = \frac{\Delta E_{\rmn{bind}}}{\Delta E_{\rmn{orb}}},
\end{equation}
where $\Delta$$E_{\rmn{bind}}$ is the difference between gravitational and thermal energy of the ejected envelope at the beginning of Roche lobe overflow (RLOF) and 
$\Delta$$E_{\rmn{orb}}$ is the orbital energy difference between the beginning and at the end of the spiralling-in process of the binary system.    

One of the main goals of this paper is to make a contribution to the studies which search for the limits to constrain the value of ``$\alpha$$_{\rmn{CE}}$" parameter. We have mainly 
followed the approximations given by, e.g., \citet{b91}, \citet{b31}, \citet{b92}, \citet{b21} in order to estimate the ``{\it efficency}" parameters for UU Sge and V477 Lyr. We assume 
that the secondary components are main--sequence stars and have not been affected by the CE interaction. For the estimation of the binding energy of the envelope 
($\Delta$$E_{\rmn{bind}}$) we have used the following equation

\begin{equation}
  {\Delta E_{\rmn{bind}}} = -\frac{GM_{\rmn{1}}M_{\rmn{env}}}{\lambda R_{\rmn{L}}},
\end{equation} 
where $M_{\rmn{1}}$ is the mass of AGB star (also assumed as the initial mass of the primary component), $M_{\rmn{env}}$ = $M_{\rmn{1}}$ -- $M_{\rmn{core}}$, is the envelope mass of AGB 
star, $M_{\rmn{core}}$ is the current mass of the primary, $R_{\rmn{L}}$ is the Roche lobe radius of the primary (at the onset of mass transfer) and $\lambda$ is a dimensionless 
parameter which depends on the detailed structure of the envelope and the stellar density distribution. According to Webbink (2007), $\lambda$ is a function of 
$m_{\rmn{e}}$$\equiv$$M_{\rmn{env}}$/$M_{\rmn{1}}$=1--($M_{\rmn{core}}$/$M_{\rmn{1}}$), and is well--approximated by

\begin{equation}
  {\lambda^{-1}} \approx 3.000-3.816{m_{\rmn{e}}}+1.041{m_{\rmn{e}}}^{2}+0.067{m_{\rmn{e}}}^{3}+0.136{m_{\rmn{e}}}^{4}.
\end{equation} 
	
	The total change in orbital energy is
\begin{equation}	
  {\Delta E_{\rmn{orb}}} = \frac{GM_{\rmn{core}}M_{\rmn{2}}}{2a_{\rmn{f}}}-\frac{G(M_{\rmn{core}}+M_{\rmn{env}})M_{\rmn{2}}}{2a_{\rmn{0}}},
\end{equation}
where $M_{\rmn{2}}$ is the mass of secondary component, $a_{\rmn{0}}$ is the initial separation of the binary system, and $a_{\rmn{f}}$ is the final orbital separation.

In order to calculate these energies one needs to estimate the initial parameters of the binary systems. As a first conjecture for the initial masses
of the primaries we have made use of the evolutionary tracks of \citet{b7}. However, by taking into consideration that 
the RLOF may occur at different stages of the binary evolution we have also used the initial/final mass relations given by \citet{b40} (also see \citealt{b41}). 
If the components experience a case--C RLOF (primary fills its Roche lobe during the He--shell burning) then the mass of the remnant
is related to initial mass of the primary by $M_{\rmn{CO}}$$\approx$0.5+0.16($M_{\rmn{1}}$ -- 2.3) if 2.3$<$$M_{\rmn{1}}$$<$4.2 (the
masses are in solar units) and by $M_{\rmn{CO}}$$\approx$0.8+0.0875($M_{\rmn{1}}$ -- 4.2) if 4.2$<$$M_{\rmn{1}}$$<$7.6. If the binary system experiences a late case--B RLOF (primary 
fills its Roche lobe shortly before the He ignition, He core is non--degenerate at this point), a degenerate CO or a ``hybrid (C+O+He)" remnant develops with an initial/final mass 
relation of $M_{\rmn{CO}}$$\approx$0.167$M_{\rmn{1}}$ -- 0.085 if 2.3$<$$M_{\rmn{1}}$$<$5, and $M_{\rmn{CO}}$$\approx$0.4+0.07$M_{\rmn{1}}$ for the range of 5$<$$M_{\rmn{1}}$$<$10. 

Considering the different stages of RLOF mentioned above, let us first investigate the case--C RLOF in which the primary components of UU Sge and V477 Lyr are CO cores. Then
the initial masses of the primary stars can be estimated as $\approx$3.1 M$_{\odot}$ and $\approx$2.35 M$_{\odot}$, respectively. If we assume that the systems have experienced a case--B 
RLOF and become degenerate CO or hybrid cores, then the initial masses of the primary components must have been around 4.2 M$_{\odot}$ and 3.5 M$_{\odot}$, respectively. By taking these 
masses as limits for the progenitors of the primaries we have performed a series of calculations to estimate the $\alpha$$_{\rmn{CE}}$ parameters of the systems (following a somewhat 
similar work to that of \citealt{b58}).

We used the theoretical AGB core mass--luminosity approximation given by Iben \& Tutukov (1984, 1985, 1993);
\begin{equation}
  {L_{\rmn{AGB}}} = 6 \times 10^{4} ({M_{\rmn{CO}}}/{M_{\odot}-0.5}),
\end{equation}
where $L_{\rmn{AGB}}$ is the maximum luminosity for a post-AGB star during the plateau phase of its evolution and also the luminosity 
at the time of RLOF. Applying this approximation, $L_{\rmn{AGB}}$ values for the current primary masses of UU Sge and V477 Lyr can be 
calculated as 7680 and 480 $L_{\odot}$, respectively.

The AGB radii of the primaries have been estimated using the AGB mass--luminosity--radius relation proposed by \citet{b36};    
\begin{equation}
  {R_{\rmn{AGB}}} = 1.125M_{1}^{-0.33}(L_{\rmn{AGB}}^{0.4}+0.383L_{\rmn{AGB}}^{0.76}),
\end{equation}
where $R_{\rmn{AGB}}$ is the AGB radius that the progenitor might have reached at the stage of RLOF. The $R_{\rmn{AGB}}$ values calculated for 
the different initial masses are also given in Table 7.

Now we can estimate the initial separations, $a_{\rmn{0}}$, of the systems with the following formula (\citealt{b28}) 
\begin{equation}
  {r_{\rmn{L}}} = \frac{0.49q^{2/3}}{0.6q^{2/3}+{\rmn{ln}}(1+q^{1/3})},
\end{equation}
where $r_{\rmn{L}}$=$R_{\rmn{L}}$/$a_{\rmn{0}}$ is the relative radius of the Roche lobe and $q$ is the mass ratio being equal to $M_{1}/M_{2}$. Adopting $R_{\rmn{L}}$=$R_{\rmn{AGB}}$ 
the initial separations have been computed for different mass ratios (see Table 7).
Equations (5) and (6) have been used for the estimation of the binding and orbital energies, then the $\alpha$$_{\rmn{CE}}$ parameters
have been determined as listed in Table 7. $P_{\rmn{i}}$, the initial orbital period of the system has been also computed using the Kepler's third law and given in Table 7. 
\begin{table*}
\centering
\caption{Possible initial configurations of the systems and the efficiency parameter, $\alpha$$_{\rmn{CE}}$.}
\begin{tabular}{@{}l|l|c|c|c|c|c|c|c@{}}
   \hline
   \hline
Object	&	$M_{1}$	&$R_{\rmn{AGB}}$ & $q$ &$a_{0}$ 	 &  $P_{\rmn{i}}$ & $\alpha$$_{\rmn{CE}}$  \\
 & $_{(M_{\odot})}$ & $_{(R_{\odot})}$ & & $_{(R_{\odot})}$&$_{(days)}$ & \\
\hline
UU Sge&   3.0  & 297 & 10.42 & 511 & 738 & 0.48 \\
& 3.1 & 294 & 10.76 & 503 & 710 &  0.51 \\
& 4.2 & 266 & 14.58 & 438 & 501 & 0.99 \\
\hline
V477 Lyr&   1.0  & 60 & 6.90 & 110 & 125 & 0.71 \\
& 2.35 & 45 & 16.21 & 73 & 47 & 4.86 \\
& 3.5 & 40 & 24.14 & 62 & 30 & 12.82\\
  \hline
V477 Lyr$^{*}$&   2.2  & 210 & 15.17 & 345 & 485 & 0.83\\
  \hline
 \end{tabular}
 \begin{list}{}{}
\item[{*}] The sdO component has been assumed as a degenerate ``He" dwarf.
 \end{list}
 \end{table*}
        
The somewhat problematic point here is that the mass of the sdO component of V477 Lyr is of the order of 0.5 M$_{\odot}$ which is a critical and controversial value that lies in the 
transition region between the masses of CO and He remnants. This situation gives us a chance to consider the sdO component as a degenerate ``He" dwarf which once have a progenitor mass 
less than 2.3 M$_{\odot}$. If we assume that the 
primary component filled its Roche lobe right after it has developed a degenerate ``He" core, then we can employ the core--mass--radius relation 
given by \citet{b40},
\begin{equation}
  {R} = 10^{3.5}M_{\rmn{He}}^{4}.
\end{equation}
Here, $R$ is the radius of primary by the time of RLOF and $M{_{\rmn{He}}}$ is the current mass of the sdO star. By adopting this radius as
$R_{\rmn{L}}$ and assuming that the initial mass of the primary was around 2.2 M$_{\odot}$ we have calculated the initial parameters and values of $\alpha$$_{\rmn{CE}}$ for the system as 
listed in Table 7.

\section{Summary and Discussion}

In this paper, we report the results of the first simultaneous two--passband light curve analysis for the binary PNNi; UU Sge and V477 Lyr. We 
have also reanalysed the published radial velocity data (PB93 and PB94) of both systems in order to increase the sensitivity of the orbital
parameters employed in the solution of the light curves. The linear limb--darkening coefficients were adjusted during the solutions of both systems. Not surprisingly, the resultant 
coefficients were found to be inconsistent with the theoretically expected ones, which was interpreted as the presence of the limb--brightening effect due to strong irradiation of the 
secondary components. In the case of UU Sge, the best solution was 
achieved at the primary temperature of 78 000 ${\rmn{K}}$ by yielding a secondary component temperature of 6136 ${\rmn{K}}$. The temperature of the primary component falls into the 
temperature range of 87 000 $\pm$ 13000 ${\rmn{K}}$ derived by BPH94. Also the temperature of secondary component indicates a good agreement with that of 6250 K estimated by BPH94.

We have determined the temperatures of the hot and cool components of V477 Lyr as 49 500 ${\rmn{K}}$ and 3874 ${\rmn{K}}$, respectively. The temperature of 
the secondary star shows a remarkable discrepancy in comparison with the temperature of 5300 K derived by PB94. We believe that having 
two--passband simultaneous light curve analysis with a different approach to the solutions yielded more realistic results in terms of the temperature
of the components of both systems. 

Nevertheless, it has to be stressed that the secondary component temperature of V477 Lyr, in contrary to that of UU Sge, seems much closer 
to the value of expected from a main--sequence star at the same mass. The discrepancy between the observed and theoretical temperatures may be 
explained in terms of the thermally relaxation stage that the secondary stars have been shown to be in. The more the mass is accumulated the 
longer the thermal--relaxation time scale is needed. During the computations, our main assumption was that the mass of the secondary stars 
has not been significantly affected by the CE interaction and the highest value for the mass of the accumulated matter was adopted as 0.05 M$_{\odot}$. However, we have to bring into 
attention that the mass of the secondary components might have undergone drastic changes due to accretion, which would also result in the increase of the luminosity (and the temperature) 
of the secondary components. 

The temperature of the primary component of V477 Lyr indicates a good agreement with the H$\beta$ Zanstra temperature of 48 000 ${\rmn{K}}$
while showing a disagreement with the He\,{\sc ii} Zanstra temperature of 82 000 ${\rmn{K}}$, those of derived by PB94. 
49 500 ${\rmn{K}}$ also lies in the range of H and He\,{\sc ii} Zanstra temperatures of 32 000 ${\rmn{K}}$ -- 60 000 ${\rmn{K}}$ determined by \citet{b43}. PB94 mentions the possible 
He\,{\sc ii} stratification in the nebula, which leads up to the variety in the He\,{\sc ii} Zanstra temperatures of the sdO component. The mass loss via thermal pulses 
may explain the likely He stratification and enhanced He abundance (PB94) in the nebula. In this case, the Zanstra temperature determined from He\,{\sc ii} nebular lines could be 
misleading, which reveals that the temperature determined by using the H nebular lines may carry more valuable information on the temperature of the sdO component. 

The comparison of the thermal time scales and the post-CE ages of the systems (Table 5) indicates that the secondary components of the binary nuclei of the PNe are still out of thermal 
equilibrium along with two other post-CE systems: HS 1136+6646 and RE 1016-053. The rest of the systems that have been reported as having oversized secondary components were found to be 
thermally relaxed relatively long ago. The oversized nature and low temperatures of these systems were interpreted as the presence of the possible magnetically active regions 
which cause the increase in the size of a low--mass main sequence star while reducing its effective temperature. Detailed spectroscopic studies 
of these systems are needed.

Finally, we have implemented several calculations to estimate the common envelope efficiency parameter, $\alpha$$_{\rmn{CE}}$, of the systems. We have determined various initial
system parameters using possible evolutionary scenarios. The possible efficiency parameters for UU Sge 
were found in the range of 0.3 $<$ $\alpha$$_{\rmn{CE}}$ $<$ 1, which is close to the range of previously suggested by theoretical studies (e.g. Han et al. 1995). Nevertheless, the 
parameter values that we have found for V477 Lyr, under the assumption of having CO or C+O+He remnant as the primary component, indicate a deviation from the theoretical findings. An 
$\alpha$$_{\rmn{CE}}$ $>$ 1  may be explained as the contribution from other sources (e.g. deposition of recombination energy into CE ejection, nuclear burning on the surface of the 
secondary star, see \citet{b41} for more detail) to the CE ejection beside the efficiency of orbital energy. On the other hand, it seems impossible to know which mechanism could have 
been such effective on the efficiency parameter. Therefore, we have found the courage of making a new assumption about the primary component of V477 Lyr. By taking into account its 
current mass of 0.508 $\pm$ 0.046 M$_{\odot}$, we assumed that the primary component is a degenerate He dwarf with an initial mass of $<$ 2.3 M$_{\odot}$. By using the mass--radius 
relation 
given by \citet{b40}, we calculated the radius of the primary at the time of RLOF (independently from its luminosity, see equation (11)).
These approaches yielded an efficiency parameter of 0.83 which indicates that about 83 percent of the orbital energy were used effectively during the ejection of the CE. 

MA would like to thank the Scientific and Technological Research Council of Turkey (T\"{U}B\.{I}TAK) and her advisor, C\.{I}, for their support during her doctoral studies. The authors 
are grateful to Howard E. Bond for his help and useful suggestions on this manuscript. We also wish to thank T\"{U}B\.{I}TAK National Observatory of Turkey (TUG) for the allocation of 
observing time. And we thank the referee very much for the comments and suggestions that have helped us to improve the manuscript. This work has been supported by the Ege University 
Research Foundation under the project number of 2002/FEN/004.

\bsp

\label{lastpage}


\begin{thebibliography}{99}
\bibitem[\protect\citeauthoryear{Abell}{1966}]{b1}Abell G. O., 1966, ApJ, 144, 259
\bibitem[\protect\citeauthoryear{Af\c{s}ar \& Bond}{2005}]{b2}Af\c{s}ar M., Bond H. E., 2005, MmSAI, 76, 608
\bibitem[\protect\citeauthoryear{Alencar \& Vaz}{1999}]{b88}Alencar S. H. P., Vaz L. P. R., 1999, A\&AS, 135, 555
\bibitem[\protect\citeauthoryear{Althaus \& Benvenuto}{1998, hereafter AB98}]{b3}Althaus L. G., Benvenuto O. G.,  1998, MNRAS, 296, 206
\bibitem[\protect\citeauthoryear{Ando et al.}{1982}]{b4}Ando H., Okazaki A., Nishimura S., 1982, PASJ, 34, 141
\bibitem[\protect\citeauthoryear{Bell et al.}{1994, hereafter BPH94}]{b5}Bell S. A., Pollacco D. L., Hilditch R. W., 1994, MNRAS, 270, 449
\bibitem[\protect\citeauthoryear{Benvenuto \& Althaus}{1998, hereafter BA98}]{b6}Benvenuto O. G., Althaus L. G., 1998, MNRAS, 293, 177
\bibitem[\protect\citeauthoryear{Bl\"{o}cker}{1995}]{b7}Bl\"{o}cker T., 1995, A\&A, 299, 755
\bibitem[\protect\citeauthoryear{Bond}{1976}]{b8}Bond H. E., 1976, PASP, 88, 192
\bibitem[\protect\citeauthoryear{Bond et al.}{1978}]{b9}Bond H. E., Liller W., Mannery E. J., 1978, ApJ, 223, 252
\bibitem[\protect\citeauthoryear{Bond}{1980}]{b10}Bond H. E., 1980, IAU Circ., 3480
\bibitem[\protect\citeauthoryear{Bond \& Grauer}{1987}]{b11}Bond H. E., Grauer A. D., 1987, in IAU Colloq. 95, Second Conference on Faint Blue Stars (Dordrecht: Kluwer), 221
\bibitem[\protect\citeauthoryear{Bond \& Livio}{1990}]{b12}Bond H. E., Livio M., 1990, ApJ, 355, 568
\bibitem[\protect\citeauthoryear{Bond}{2000}]{b90}Bond H. E., 2000, in Kastner J. H., Soker N., Rappaport S., eds, ASP Conf.
Ser. Vol. 199, Asymmetrical Planetary Nebulae II: From Origins to Microstructures.
Astron. Soc. Pac., San Francisco, p. 115
\bibitem[\protect\citeauthoryear{Bruch \& Diaz}{1998}]{b13}Bruch A., Diaz M. P., 1998, AJ, 116, 908
\bibitem[\protect\citeauthoryear{Bruch}{1999}]{b14}Bruch A. 1999, AJ, 117, 3031
\bibitem[\protect\citeauthoryear{Bruch \& Diaz}{1999}]{b15}Bruch A., Diaz M. P., 1999, A\&A, 351, 573
\bibitem[\protect\citeauthoryear{Budding \& Kopal}{1980}]{b16}Budding E., Kopal Z., 1980, Ap\&SS, 73, 83
\bibitem[\protect\citeauthoryear{Caillault \& Patterson}{1990}]{b17}Caillault A. P., Patterson J., 1990, AJ, 100, 825
\bibitem[\protect\citeauthoryear{Chabrier \& Baraffe}{1997, hereafter CB97}]{b18}Chabrier G., Baraffe I., 1997, A\&A, 327, 1039
\bibitem[\protect\citeauthoryear{Ciardullo et al.}{1999}]{b19}Ciardullo R., Bond H. E., Sipior M. S., Fullton L. K., Zhang C.-Y., Schaefer K. G., 1999, AJ, 118, 488
\bibitem[\protect\citeauthoryear{Claret}{1998}]{b20}Claret A., 1998, A\&A, 335, 647
\bibitem[\protect\citeauthoryear{Claret et al.}{1995}]{b84}Claret A., D\'{\i}az-Cordov\'{e}s J., Gim\'{e}nez A., 1995, A\&AS, 114, 247
\bibitem[\protect\citeauthoryear{Claret}{2004}]{b86}Claret A., 2004, A\&A, 422, 665
\bibitem[\protect\citeauthoryear{De Kool}{1990}]{b21}De Kool M., 1990, ApJ, 358, 189
\bibitem[\protect\citeauthoryear{De Marco et al.}{2004}]{b22}De Marco O., Bond H. E., Harmer D., Fleming A. J., 2004, ApJ, 602, L93
\bibitem[\protect\citeauthoryear{De Marco et al.}{2008}]{b23}De Marco O., Hillwig, T. C., Smith A. J., 2008, AJ, 136, 323
\bibitem[\protect\citeauthoryear{Dewi \& Tauris}{2000}]{b92}Dewi J. D. M., Tauris T. M., 2000, A\&A, 360, 1043
\bibitem[\protect\citeauthoryear{D\'{\i}az-Cordov\'{e}s et al.}{1995}]{b24}D\'{\i}az-Cordov\'{e}s J., Claret A., Gim\'{e}nez A., 1995, A\&AS, 110, 329
\bibitem[\protect\citeauthoryear{Drilling}{1983}]{b26}Drilling J. S., 1983, ApJ, 270, L13
\bibitem[\protect\citeauthoryear{Drilling}{1985}]{b27}Drilling J. S., 1985, ApJ, 294, 107
\bibitem[\protect\citeauthoryear{Eggleton}{1983}]{b28}Eggleton P. P., 1983, ApJ, 268, 368
\bibitem[\protect\citeauthoryear{Exter et al.}{2005}]{b89}Exter K. M., Pollacco D. L., Maxted P. F. L., Napiwotzki R., Bell S. A., 2005, MNRAS, 359, 315
\bibitem[\protect\citeauthoryear{Ferguson et al.}{1999}]{b29}Ferguson D. H., Liebert J., Haas S., Napiwotzki R., James T. A., 1999, ApJ, 518, 866
\bibitem[\protect\citeauthoryear{Green et al.}{1986}]{b30}Green R. F., Schmidt M., Liebert J., 1986, ApJS, 61, 305
\bibitem[\protect\citeauthoryear{Han et al.}{1995}]{b31}Han Z., Podsiadlowski P., Eggleton P. P., 1995, MNRAS, 272, 800
\bibitem[\protect\citeauthoryear{Heber et al.}{1996}]{b32}Heber U., Dreizler S., Hagen H. J., 1996, A\&A, 311, L17
\bibitem[\protect\citeauthoryear{Hilditch et al.}{1996}]{b33}Hilditch R. W., Harries T. J., Hill G., 1996, MNRAS, 279, 1380
\bibitem[\protect\citeauthoryear{Hjellming \& Taam}{1991}]{b34}Hjellming M. S., Taam, R. E., 1991, ApJ, 370, 709
\bibitem[\protect\citeauthoryear{Hoffleit}{1932}]{b35}Hoffleit, D., 1932, Harvard Bull., No. 887
\bibitem[\protect\citeauthoryear{Hurley et al.}{2000}]{b36}Hurley J. R., Pols O. R., Tout C. A., 2000, MNRAS, 315, 543
\bibitem[\protect\citeauthoryear{\.{I}bano\v{g}lu et al.}{1994}]{b37}\.{I}bano\v{g}lu C., Keskin V., Akan M. C., Evren S., Tunca Z., 1994, A\&A, 281, 811
\bibitem[\protect\citeauthoryear{Iben \& Tutukov}{1984}]{b38}Iben I. Jr., Tutukov A. V., 1984, ApJS, 54, 335
\bibitem[\protect\citeauthoryear{Iben \& Tutukov}{1985}]{b39}Iben I. Jr., Tutukov A. V., 1985, ApJS, 58, 661
\bibitem[\protect\citeauthoryear{Iben \& Tutukov}{1986}]{b40}Iben I. Jr., Tutukov A. V., 1986, ApJ, 311, 753
\bibitem[\protect\citeauthoryear{Iben \& Livio}{1993}]{b41}Iben I. Jr., Livio M., 1993, PASP, 105, 1373
\bibitem[\protect\citeauthoryear{Iben \& Tutukov}{1993}]{b42}Iben I. Jr., Tutukov A. V., 1993, ApJ, 418, 343
\bibitem[\protect\citeauthoryear{Kaler}{1983}]{b43}Kaler J. B., 1983, ApJ, 271, 188
\bibitem[\protect\citeauthoryear{Kiss et al.}{2000}]{b44}Kiss L. L., Kasza J., Borza S., 2000, IBVS, 4962
\bibitem[\protect\citeauthoryear{Krzeminski}{1984}]{b45}Krzeminski W., 1984, IAU Circ. 4014
\bibitem[\protect\citeauthoryear{Kurochkin}{1964}]{b46}Kurochkin N. E., 1964, Peremnye Zvezdy (Academy Sci. U.S.S.R., Var. Star Bull), 15, 77
\bibitem[\protect\citeauthoryear{Kurucz}{1979}]{b47}Kurucz, R. L., 1979, ApJS, 40, 1
\bibitem[\protect\citeauthoryear{Kurucz}{1991}]{b48}Kurucz R. L., 1991, in Davis Philips A. C., Upgren A. R., James K. A., eds, Proc. Workshop on Precision Photometry: Astrophysics of 
the Galaxy. Schenectady, NY, p. 27
\bibitem[\protect\citeauthoryear{Kurucz}{1993}]{b49}Kurucz R. L., 1993, in Milone E. F., ed., Light Curve Modelling of Eclipsing Binary Stars. Springer-Verlag, New York, p. 93
\bibitem[\protect\citeauthoryear{Lacy}{1977}]{b50}Lacy C. H., 1977, ApJS, 34, 479
\bibitem[\protect\citeauthoryear{Leggett}{1992}]{b51}Leggett S. K., 1992, ApJS, 82, 351
\bibitem[\protect\citeauthoryear{Liebert et al.}{1995}]{b52}Liebert J., Tweedy R., Napiwotzki R., Fulbright M. S., 1995, ApJ, 441, 424
\bibitem[\protect\citeauthoryear{Maxted et al.}{2007}]{b54}Maxted P. F. L., OíDonoghue D., Morales-Rueda L., Napiwotzki R., Smalley B., 2007, MNRAS, 376, 919
\bibitem[\protect\citeauthoryear{Morales et al.}{2008}]{b55}Morales J. C., Ribas I., Jordi C., 2008, A\&A, 478, 507
\bibitem[\protect\citeauthoryear{Mullan \& MacDonald}{2001}]{b56}Mullan D. J., MacDonald J., 2001, ApJ, 559, 353
\bibitem[\protect\citeauthoryear{Nelson \& Young}{1970}]{b57}Nelson B., Young A., 1970, PASP, 82, 699
\bibitem[\protect\citeauthoryear{O'Brien et al.}{2001}]{b58}O'Brien M. S., Bond H. E., Sion E. M., 2001, ApJ, 563, 971
\bibitem[\protect\citeauthoryear{Paczy\`{n}ski}{1976}]{b59}Paczy\`{n}ski B., 1976, in IAU Symp. 73, Structure and Evolution of Close Binary Systems, ed. P. Eggleton, S. Mitton, J. Whelan 
(Dordrecht: Reidel), 75
\bibitem[\protect\citeauthoryear{Peraiah}{1983}]{b87}Peraiah A., 1983, JApA, 4, 11
\bibitem[\protect\citeauthoryear{Pollacco \& Bell}{1993, hereafter PB93}]{b60}Pollacco D. L., Bell S. A., 1993, MNRAS, 262, 377
\bibitem[\protect\citeauthoryear{Pollacco \& Bell}{1994, hereafter PB94}]{b61}Pollacco D. L., Bell S. A., 1994, MNRAS, 267, 452
\bibitem[\protect\citeauthoryear{Prialnik \& Livio}{1985}]{b62}Prialnik D., Livio M., 1985, MNRAS, 216, 37
\bibitem[\protect\citeauthoryear{Ritter}{1986}]{b63}Ritter H., 1986, A\&A, 169, 139
\bibitem[\protect\citeauthoryear{Saffer et al.}{1993}]{b64}Saffer R. A., Wade R. A., Liebert J., Green R. F., Sion E. M., Bechtold J., Foss D., Kidder K., 1993, AJ, 105, 1945
\bibitem[\protect\citeauthoryear{Sandquist et al.}{1998}]{b65}Sandquist E. L., Taam R. E., Chen X., Bodenheimer P., Burkert A., 1998, ApJ, 500, 909
\bibitem[\protect\citeauthoryear{Sarna \& Zi\'{o}{\l}kowski}{1988}]{b66}Sarna M. J., Zi\'{o}{\l}kowski J., 1988, AcA, 38, 89
\bibitem[\protect\citeauthoryear{Sch\"{o}nberner}{1981}]{b67}Sch\"{o}nberner D., 1981, A\&A, 103, 119
\bibitem[\protect\citeauthoryear{Sch\"{o}nberner}{1983}]{b68}Sch\"{o}nberner D., 1983, ApJ, 272, 708
\bibitem[\protect\citeauthoryear{Sch\"{o}nberner \& Drilling}{1984}]{b69}Sch\"{o}nberner D., Drilling J. S., 1984, ApJ, 278, 702
\bibitem[\protect\citeauthoryear{Sing et al.}{2004}]{b70}Sing D. K., Holberg J. B., Burleigh M. R., Good S. A., Barstow M. A., Oswalt T. D., Howell S. B., Brinkworth C. S. et al., 2004, 
AJ, 127, 2936
\bibitem[\protect\citeauthoryear{Soker}{2006}]{b71}Soker N., 2006, ApJ, 645L, 57
\bibitem[\protect\citeauthoryear{Tutukov \& Yungelson}{1979}]{b73}Tutukov A. V., Yungelson L. R., 1979, in Mass loss and evolution of O-type stars, ed. C. de Loore, P. S. Conti (Reidel, 
Dordrecht) 401
\bibitem[\protect\citeauthoryear{Van Buren et al.}{1980}]{b74}Van Buren D., Charles P. A., Mason K. O., 1980, ApJ, 242L, 105
\bibitem[\protect\citeauthoryear{Van Hamme}{1993, note that the limb-darkening coefficients given in these tables are calculated for up to an effective temperature of 50 000 K, therefore 
in this paper, the bolometric limb-darkening coefficient for the primary component was always kept at a value that corresponds to 50 000 K}]{b75}Van Hamme W., 1993, AJ, 106, 2096
\bibitem[\protect\citeauthoryear{Vennes et al.}{1999}]{b76}Vennes S., Thorstensen J. R., Polomski E. F., 1999, ApJ, 523, 386
\bibitem[\protect\citeauthoryear{Walton et al.}{1993}]{b77}Walton N. A., Walsh J. R., Pottasch S. R., 1993, A\&A, 275, 256 
\bibitem[\protect\citeauthoryear{Webbink}{2008}]{b91}Webbink R. F., 2008, Short-Period Binary Stars: Observations, Analyses, and Results. Series: Astrophysics and Space Science Library, 
XVIII, Vol. 352, p.233. Edited by E.F. Milone, D.A. Leahy, and D.W. Hobill
\bibitem[\protect\citeauthoryear{Wilson}{1979}]{b78}Wilson R. E., 1979, ApJ, 234, 1054
\bibitem[\protect\citeauthoryear{Wilson}{1990}]{b79}Wilson R. E., 1990, ApJ, 356, 613
\bibitem[\protect\citeauthoryear{Wilson \& Devinney}{1971}]{b80}Wilson R. E., Devinney E. J., 1971, ApJ, 166, 605
\bibitem[\protect\citeauthoryear{Wilson \& Van Hamme}{2003}]{b81}Wilson R. E., Van Hamme W., 2003, Documentation of Eclipsing Binary Computer Model.
Dept. Astron., Univ. Florida, Gainesville, FL, USA
\bibitem[\protect\citeauthoryear{Wood et al.}{1993}]{b85}Wood J. H., Zhang E-H., Robinson E. L., 1993, MNRAS, 261, 103
\bibitem[\protect\citeauthoryear{Wood et al.}{1999}]{b82}Wood J. H., Harmer S., Lockley J. J., 1999, MNRAS, 304, 335
\end{thebibliography}
\end{document}